\documentclass[prc,twocolumn]{revtex4-1}

\usepackage{enumerate}

\usepackage{booktabs}
\usepackage{color}
\usepackage{graphicx}

\usepackage{amsmath}
\usepackage{amssymb}
\usepackage{times}

\def\bvec#1{\mbox{\boldmath $#1$}}

\newcommand{\Tr}[0]{\mbox{Tr}}

\newcommand{\del}[2]{\frac{\partial #1}{\partial #2}}

\newcommand{\bra}{\langle}
\newcommand{\ket}{\rangle}

\newcommand{\idot}{\cdot}

\newcommand{\beq}{\begin{equation}}
\newcommand{\eeq}{\end{equation}}
\newcommand{\bea}{\begin{eqnarray}}
\newcommand{\eea}{\end{eqnarray}}

\def\fun#1#2{\lower3.6pt\vbox{\baselineskip0pt\lineskip.9pt
 \ialign{$\mathsurround=0pt#1\hfil##\hfil$\crcr#2\crcr\sim\crcr}}}

\begin{document}

\title{
  Complex eigenenergy of GDR for $^{16}$O by the Jost function within RPA framework
}

\author{K. Mizuyama$^{1,2}$, N. Nhu Le$^{3,4}$, T. V. Nhan Hao$^{3,4}$, N. Hoang Tung$^{5}$, and
  T. Dieu Thuy$^{3}$}
\email{corresponding author: tthuy1406@gmail.com}

\affiliation{
  \textsuperscript{1}
  Institute of Research and Development, Duy Tan University,
  Da Nang 550000, Vietnam
  \\
  \textsuperscript{2}
  Faculty of Natural Sciences,  Duy Tan University, Da Nang 550000, Vietnam
  \\
  \textsuperscript{3}
  Faculty of Physics, University of Education, Hue University,
  34 Le Loi Street, Hue City, Vietnam
  \\
  \textsuperscript{4} Center for Theoretical and Computational Physics,
  College of Education, Hue University, 34 Le Loi Street, Hue City, Vietnam
  \\
  \textsuperscript{5} Faculty of Physics, The University of Danang,
  University of Science and Education, Da Nang 550000, Vietnam
}

\date{\today}

\begin{abstract}
  The Jost function method is extended within the framework of the RPA theory to find poles
  on the complex energy plane that exhibit complex RPA eigenenergies.
  Poles corresponding to the RPA excited states such as the giant resonance of $^{16}$O electric
  dipole excitations were successfully found on the complex energy plane.
  Although the giant resonance has been known as a single resonance with large strength and
  width, it is found that, at least withing the RPA framework, the $^{16}$O electric dipole
  giant resonance is formed by multiple
  poles, each of which is an independent pole with different widths, origins, response
  properties to residual interactions, and components structures of the density fluctuation..
\end{abstract}

\maketitle

\section{Introduction}

In nuclear physics, the giant resonance is known as a resonance with
a very broad and large strength that appears in the relatively higher continuum energy region
above threshold
in the cross section or strength function expressed as a function of the excitation
energy~\cite{baldwin,berman,harakeh,bortignon}. 
The random-phase-approximation (RPA)~\cite{bohm} is a powerful tool for the description of
the collective excitation of nuclei, such as a giant resonance~\cite{liu,ring-speth}. 
Based on the understanding of the schematic model of RPA~\cite{ring},
the RPA solution that is
formed by the superposition of a large number of $p$-$h$ excited configurations and gives a large
strength due to the effect of residual interactions is interpreted as a collective excitation
mode, and the giant resonance is considered to be such a collective excitation mode.

There are two main methods for solving the equation in RPA theory:
one that uses a discrete basis and diagonalises the RPA
Hamiltonian to obtain a solution, and the continuum RPA (cRPA)~\cite{shlomo}
that takes into account the boundary conditions of the continuum.
In the former method, all RPA solutions are obtained as
discrete energy eigenstates, with giant resonances appearing either as solutions giving large
strength at the resonance energy or as many spread discrete solutions near the resonance energy.
The width is often evaluated from the empirical reproducibility of the experimental data
using the Lorentz distribution function, assuming that the giant resonance has a peak structure
with a single width.
Since the latter method (the cRPA method) takes into account the boundary conditions of the
continuum, it can represent the behavior of the strength function as a smooth function
of energy above threshold.
Without the assumption of a Lorentz distribution, as a result of numerical calculations,
the giant resonance is also represented as a smooth function of energy, which has a peak
structure with a width. 
However, the cRPA method is not able to calculate the width itself.

There is a phenomenological model that describes the width of the giant resonance in terms of
mass number and dissipative term.
It is well known that the mass number dependence of experimental values of the giant resonance
width estimated by the Lorenz distribution function roughly follows this phenomenological
model~\cite{speth,aue}.
This phenomenological model is derived from the Euler equations plus a dissipation term
representing viscosity, assuming that the giant resonance is a harmonic oscillation with damping
due to viscosity. 
In calculations such as the second RPA, which include higher-order effects, the discrete
strength (of resonances such as the giant resonance) obtained from calculations using a
discrete basis is fragmented and spread in energy by the higher-order effects. 
This spreading of the strength is interpreted as the broadening of the width of the resonance
due to higher-order effects (spreading width), however, this is not an argument made by directly
calculating the width.~\cite{bortignon}. 
Despite the fact that width is one of the most important characters in the giant resonance,
there is still no method that can calculate the width of resonances obtained as the complex
eigenenergy states of the nucleus within the microscopic theoretical framework such as
the RPA theory.

The Jost function is a function that gives the energy eigenvalues of the fundamental
differential equations of a quantum system based on the Hamiltonian of a system such
as the Schr\"{o}dinger equation as zeros on the complex energy plane.
The Jost function is given in differential and integral forms, and its differential form is
equivalent to the Wronskian. The energy eigenvalues given by the zeros on the complex energy
plane give not only bound states but also resonance states.
The imaginary part of the zeros on the complex energy plane of the Jost function gives the width
of the resonance. 
However, the original Jost function~\cite{jost} did not take into account channel coupling
(Note that the term ``channel coupling'' here includes a wide range of meanings, such as
the coupling
between $p$-$h$ configurations). 
In order to apply the Jost function to solve various physics problems, it is necessary to
extend the Jost function to be able to consider coupled channels.
As a first step, we have extended the Jost function to the Hartree-Fock-Bogoliubov theory 
framework in Refs.\cite{jost-hfb,jost-fano,jost-class}.

In this paper, we extend the Jost function method within the framework of RPA theory in order to
enable the Jost function method to find the complex eigenenergies of the RPA solutions on the
complex energy plane.
The electric dipole
excitation of $^{16}$O is then chosen as the first application of the Jost-RPA method and analyzed
by calculating the poles of the RPA strength function on the complex energy plane, adopting the
Woods-Saxon potential for the mean field, and the simple density dependent contact interaction
for the residual interaction. 

\section{Formalism of Jost function for RPA}
\label{sec2}
Since the Jost function is defined as a coefficient function relating regular and irregular
solutions of second-order differential equations such as the Schr\"{o}dinger equation, it is
necessary to represent the RPA equation in the form of a second-order differential equation
in order to define and calculate the Jost function within the framework of RPA theory. 

In this section, we present the derivation of the Jost function within the framework of RPA
theory and the perturbed Green's function, RPA response function, and strength function using
the Jost function.

\subsection{Derivation of the Jost function}

Defining the functions $X_h(\bvec{r})$ and $Y_h(\bvec{r})$ as
\begin{eqnarray}
  X_h(\bvec{r};\omega)
  &=&
  \sum_p
  X_{ph}(\omega)
  \varphi_p(\bvec{r})
  \\
  Y_h(\bvec{r};\omega)
  &=&
  \sum_p
  Y_{ph}(\omega)
  \varphi_p^*(\bvec{r})
\end{eqnarray}
using the $X$ and $Y$ amplitudes in the ordinary RPA equation known
to be expressed in the form
$\begin{pmatrix}
  A & B \\
  B & A
\end{pmatrix}
\begin{pmatrix}
  X \\ Y
\end{pmatrix}
=
\hbar\omega
\begin{pmatrix}
  1 & 0 \\
  0 & -1
\end{pmatrix}
\begin{pmatrix}
  X \\ Y
\end{pmatrix}$
and the single particle wave function $\varphi_p$, 
the differential equation for the partial wave component of $X_h(\bvec{r})$ and $Y_h(\bvec{r})$ 
in a spherically symmetric system is given by
\begin{eqnarray}
  &&
  \begin{pmatrix}
    h_{lj}^{(q)}-\epsilon_{h}^{(q)}-\omega
    &
    0
    \\
    0
    &
    h_{lj}^{(q)}-\epsilon_{h}^{(q)}+\omega
  \end{pmatrix}
  \begin{pmatrix}
    X_{lj;h}^{L(q)} \\
    Y_{lj;h}^{L(q)}
  \end{pmatrix}
  \nonumber\\
  &&
  +
  \sum_{q'h'l'j'}
  \frac{\kappa_{qq'}(r)}{r^2}
  \begin{pmatrix}
    \tilde{\varphi}_{lj;h}^{L(q)}
    \tilde{\varphi}_{l'j';h'}^{L(q')}
    &
    \tilde{\varphi}_{lj;h}^{L(q)}
    \tilde{\varphi}_{l'j';h'}^{L(q')}
    \\
    \tilde{\varphi}_{lj;h}^{L(q)}
    \tilde{\varphi}_{l'j';h'}^{L(q')}
    &
    \tilde{\varphi}_{lj;h}^{L(q)}
    \tilde{\varphi}_{l'j';h'}^{L(q')}
  \end{pmatrix}
  \begin{pmatrix}
    X_{l'j';h'}^{L(q')} \\
    Y_{l'j';h'}^{L(q')}
  \end{pmatrix}
  \nonumber\\
  &&=0
  \label{RPAeq0}
\end{eqnarray}
where $h_{lj}^{(q)}$ is the mean field Hamiltonian given by
\begin{eqnarray}
  h_{lj}^{(q)}
  =
  -\frac{\hbar^2}{2m}
  \del{^2}{r^2}
  +
  U_{lj}^{(q)}(r)
\end{eqnarray}
(the centrifugal potential $\frac{\hbar^2l(l+1)}{2mr^2}$ is included in $U_{lj}^{(q)}(r)$),  
and $\kappa_{qq'}(r)$ is the residual interaction, and $\tilde{\varphi}_{lj;h}^{L(q)}$
is the function which is defined by 
\begin{eqnarray}
  \tilde{\varphi}_{lj;h}^L(r)
  \equiv
  \frac{\bra lj||Y_L||l_{h}j_{h}\ket}{\sqrt{2L+1}}
  \varphi_{h}^{(q)}(r)
\end{eqnarray}
with use of the hole state wave function $\varphi_{h}^{(q)}(r)$ satisfies
$h_{l_hj_h}^{(q)}\varphi_{h}^{(q)}=e_h^{(q)}\varphi_{h}^{(q)}$. The subscription
$h$ describes a hole state quantum numbers as $h\in(n_h,l_h,j_h)$, and
$q$ denotes neutron or proton by $q=n$ or $p$.

By introducing a subscription $\alpha$ which expresses the particle-hole
transition configuration (as shown in Table.\ref{table1}), 
and the momentum $k_{1,\alpha}^{(q)}$ and $k_{2,\alpha}^{(q)}$ which are defined by 
\begin{eqnarray}
  k_{1,\alpha}^{(q)}(\omega)
  &=&
  \sqrt{\frac{2m}{\hbar^2}(e^{(q)}_{\alpha}+\omega)}
  \label{k1def}
  \\
  k_{2,\alpha}^{(q)}(\omega)
  &=&
  \sqrt{\frac{2m}{\hbar^2}(e^{(q)}_{\alpha}-\omega)},
  \label{k2def}
\end{eqnarray}
Eq.(\ref{RPAeq0}) can be rewritten as
\begin{eqnarray}
  &&
  \sum_{q'\alpha'}
  \left[
  \frac{\hbar^2}{2m}
    \begin{pmatrix}
      k_{1,\alpha}^{(q)2}
      &
      0
      \\
      0
      &
      k_{2,\alpha}^{(q)2}
    \end{pmatrix}
    \delta_{qq'}\delta_{\alpha\alpha'}
    \right.
  \nonumber\\
  &&
  \left.
  -
  \left\{
    -\frac{\hbar^2}{2m}
    \del{^2}{r^2}
    \begin{pmatrix}
      1
      &
      0
      \\
      0
      &
      1
    \end{pmatrix}
    \delta_{qq'}\delta_{\alpha\alpha'}
    \right.
    \right.
  \nonumber\\
  &&
  \left.
  \left.
  +
  U_{\alpha}^{(q)}
  \begin{pmatrix}
      1
      &
      0
      \\
      0
      &
      1
  \end{pmatrix}
  \delta_{qq'}\delta_{\alpha\alpha'}
  \right.
  \right.
  \nonumber\\
  &&
  \left.
  \left.
  +
  \frac{\kappa_{qq'}(r)}{r^2}
  \begin{pmatrix}
    \tilde{\varphi}_{\alpha}^{(q)}
    \tilde{\varphi}_{\alpha'}^{(q')}
    &
    \tilde{\varphi}_{\alpha}^{(q)}
    \tilde{\varphi}_{\alpha'}^{(q')}
    \\
    \tilde{\varphi}_{\alpha}^{(q)}
    \tilde{\varphi}_{\alpha'}^{(q')}
    &
    \tilde{\varphi}_{\alpha}^{(q)}
    \tilde{\varphi}_{\alpha'}^{(q')}
  \end{pmatrix}
  \right\}
  \right]
  \begin{pmatrix}
    X_{\alpha'}^{(q')} \\
    Y_{\alpha'}^{(q')}
  \end{pmatrix}
  \nonumber\\
  &&=0.
  \label{RPAeq1}
\end{eqnarray}
Since $\alpha$ is defined for each multipolarity $L$ (i.e., the multipolarity $L$ is fixed
when $\alpha$ is defined), thereafter we will not explicitly show $L$ in the formula.

Furthermore, Eq.(\ref{RPAeq1}) can be represented in the matrix form as
\begin{eqnarray}
  \left[
    \frac{\hbar^2}{2m}
    \bvec{\mathcal{K}}^2
    -    
    \left\{
    -
    \frac{\hbar^2}{2m}
    \del{^2}{r^2}
    \bvec{1}
    +
    \bvec{\mathcal{U}}
    +
    \bvec{\mathcal{V}}
    \right\}
    \right]
  \vec{\bvec{\phi}}
  =0
  \label{RPAeq2}
\end{eqnarray}
where $\bvec{\mathcal{K}}$ and $\bvec{\mathcal{U}}$ are
defined as the $N_n+N_p$ dimensional diagonal matrix defined as
\begin{eqnarray}
  \bvec{\mathcal{K}}
  =
  \begin{pmatrix}
    \begin{matrix}
      \mathcal{K}_{1}^{(n)} & 0 & \cdots & 0 \\
      0 & \mathcal{K}_{2}^{(n)} & \cdots & 0 \\
      \vdots & \vdots &\ddots & \vdots  \\
      0 & 0 & \cdots & \mathcal{K}_{N_n}^{(n)}
    \end{matrix}
    &
    \begin{matrix}
      &&& \\
      &\text{\huge{0}}&& \\
      &&& \\
      &&& \\
    \end{matrix}
    \\
    \begin{matrix}
      &&& \\
      &\text{\huge{0}}&& \\
      &&& \\
      &&& \\
    \end{matrix}
    &
    \begin{matrix}
      \mathcal{K}_{1}^{(p)} & 0 & \cdots & 0 \\
      0 & \mathcal{K}_{2}^{(p)} & \cdots & 0 \\
      \vdots & \vdots &\ddots & \vdots  \\
      0 & 0 & \cdots & \mathcal{K}_{N_p}^{(p)}
    \end{matrix}
  \end{pmatrix}
\end{eqnarray}
and
\begin{eqnarray}
  \bvec{\mathcal{U}}
  =
  \begin{pmatrix}
    \begin{matrix}
      \mathcal{U}_{1}^{(n)} & 0 & \cdots & 0 \\
      0 & \mathcal{U}_{2}^{(n)} & \cdots & 0 \\
      \vdots & \vdots &\ddots & \vdots  \\
      0 & 0 & \cdots & \mathcal{U}_{N_n}^{(n)}
    \end{matrix}
    &
    \begin{matrix}
      &&& \\
      &\text{\huge{0}}&& \\
      &&& \\
      &&& \\
    \end{matrix}
    \\
    \begin{matrix}
      &&& \\
      &\text{\huge{0}}&& \\
      &&& \\
      &&& \\
    \end{matrix}
    &
    \begin{matrix}
      \mathcal{U}_{1}^{(p)} & 0 & \cdots & 0 \\
      0 & \mathcal{U}_{2}^{(p)} & \cdots & 0 \\
      \vdots & \vdots &\ddots & \vdots  \\
      0 & 0 & \cdots & \mathcal{U}_{N_p}^{(p)}
    \end{matrix}
  \end{pmatrix}
\end{eqnarray}
by using the $2\times 2$ matrices
\begin{eqnarray}
  \mathcal{K}_{\alpha}^{(q)}
  =
  \begin{pmatrix}
      k_{1,\alpha}^{(q)}
      &
      0
      \\
      0
      &
      k_{2,\alpha}^{(q)}
  \end{pmatrix}
\end{eqnarray}
and
\begin{eqnarray}
  \mathcal{U}_{\alpha}^{(q)}
  =
  U_{\alpha}^{(q)}
  \begin{pmatrix}
      1
      &
      0
      \\
      0
      &
      1
  \end{pmatrix},
\end{eqnarray}
therefore $\bvec{\mathcal{K}}$ and $\bvec{\mathcal{U}}$ are
totally $2(N_n+N_p)\times 2(N_n+N_p)$ dimensional diagonal matrix. $N_q$(for $q=n$ and $p$)
is the number of the p-h configurations. (For example,
$N_n=N_p=7$ in the case of the electric dipole excitation of $^{16}$O
as shown in Table.\ref{table1}).

$\bvec{\mathcal{V}}$ is the $2(N_n+N_p)\times 2(N_n+N_p)$ matrix
for the residual interaction which is represented as
\begin{eqnarray}
  &&
  \bvec{\mathcal{V}}
  =
  \nonumber\\
  &&
  \begin{pmatrix}
    \begin{matrix}
      \mathcal{V}_{11}^{(nn)} &
      \mathcal{V}_{12}^{(nn)} &
      \cdots &
      \mathcal{V}_{1N_n}^{(nn)}
      \\
      \mathcal{V}_{21}^{(nn)} &
      \mathcal{V}_{22}^{(nn)} &
      \cdots &
      \mathcal{V}_{2N_n}^{(nn)}
      \\
      \vdots &
      \vdots &
      \ddots &
      \vdots
      \\
      \mathcal{V}_{N_n1}^{(nn)} &
      \mathcal{V}_{N_n2}^{(nn)} &
      \cdots &
      \mathcal{V}_{N_nN_n}^{(nn)}
    \end{matrix}
    &
    \begin{matrix}
      \mathcal{V}_{11}^{(np)} &
      \mathcal{V}_{12}^{(np)} &
      \cdots &
      \mathcal{V}_{1N_p}^{(np)}
      \\
      \mathcal{V}_{21}^{(np)} &
      \mathcal{V}_{22}^{(np)} &
      \cdots &
      \mathcal{V}_{2N_p}^{(np)}
      \\
      \vdots &
      \vdots &
      \ddots &
      \vdots
      \\
      \mathcal{V}_{N_n1}^{(np)} &
      \mathcal{V}_{N_n2}^{(np)} &
      \cdots &
      \mathcal{V}_{N_nN_p}^{(np)}
    \end{matrix}
    \\
    \begin{matrix}
      \mathcal{V}_{11}^{(pn)} &
      \mathcal{V}_{12}^{(pn)} &
      \cdots &
      \mathcal{V}_{1N_n}^{(pn)}
      \\
      \mathcal{V}_{21}^{(pn)} &
      \mathcal{V}_{22}^{(pn)} &
      \cdots &
      \mathcal{V}_{2N_n}^{(pn)}
      \\
      \vdots &
      \vdots &
      \ddots &
      \vdots
      \\
      \mathcal{V}_{N_p1}^{(pn)} &
      \mathcal{V}_{N_p2}^{(pn)} &
      \cdots &
      \mathcal{V}_{N_pN_n}^{(pn)}
    \end{matrix}
    &
    \begin{matrix}
      \mathcal{V}_{11}^{(pp)} &
      \mathcal{V}_{12}^{(pp)} &
      \cdots &
      \mathcal{V}_{1N_p}^{(pp)}
      \\
      \mathcal{V}_{21}^{(pp)} &
      \mathcal{V}_{22}^{(pp)} &
      \cdots &
      \mathcal{V}_{2N_p}^{(pp)}
      \\
      \vdots &
      \vdots &
      \ddots &
      \vdots
      \\
      \mathcal{V}_{N_p1}^{(pp)} &
      \mathcal{V}_{N_p2}^{(pp)} &
      \cdots &
      \mathcal{V}_{N_pN_p}^{(pp)}
    \end{matrix}
  \end{pmatrix}
  \nonumber\\
  \label{Vres-mat}
\end{eqnarray}
with 
\begin{eqnarray}
  \mathcal{V}_{\alpha\alpha'}^{(qq')}
  =
  \frac{\kappa_{qq'}(r)}{r^2}
  \begin{pmatrix}
    \tilde{\varphi}_{\alpha}^{(q)}
    \tilde{\varphi}_{\alpha'}^{(q')}
    &
    \tilde{\varphi}_{\alpha}^{(q)}
    \tilde{\varphi}_{\alpha'}^{(q')}
    \\
    \tilde{\varphi}_{\alpha}^{(q)}
    \tilde{\varphi}_{\alpha'}^{(q')}
    &
    \tilde{\varphi}_{\alpha}^{(q)}
    \tilde{\varphi}_{\alpha'}^{(q')}.
  \end{pmatrix}
\end{eqnarray}
Note that $\bvec{\mathcal{V}}$ is a symmetric matrix.
If we define $2(N_n+N_p)$-dimensional vector for the hole state wave functions
$\tilde{\varphi}_{\alpha}^{(q)}$ as
\begin{eqnarray}
  \vec{\tilde{\bvec{\varphi}}}_n
  =
  \begin{pmatrix}
    \begin{pmatrix}
      \tilde{\varphi}_{1}^{(n)} \\
      \tilde{\varphi}_{1}^{(n)}
    \end{pmatrix}
    \\
    \begin{pmatrix}
      \tilde{\varphi}_{2}^{(n)} \\
      \tilde{\varphi}_{2}^{(n)}
    \end{pmatrix}\\
    \vdots \\
    \begin{pmatrix}
      \tilde{\varphi}_{N_n}^{(n)} \\
      \tilde{\varphi}_{N_n}^{(n)}
    \end{pmatrix}
    \\
    \begin{pmatrix}
      0\\
      0
    \end{pmatrix}
    \\
    \begin{pmatrix}
      0\\
      0
    \end{pmatrix}
    \\
    \vdots \\
    \begin{pmatrix}
      0\\
      0
    \end{pmatrix}
  \end{pmatrix}
  \mbox{, and }
  \vec{\tilde{\bvec{\varphi}}}_p
  =
  \begin{pmatrix}
    \begin{pmatrix}
      0\\
      0
    \end{pmatrix}
    \\
    \begin{pmatrix}
      0\\
      0
    \end{pmatrix}
    \\
    \vdots \\
    \begin{pmatrix}
      0\\
      0
    \end{pmatrix}
    \\
    \begin{pmatrix}
      \tilde{\varphi}_{1}^{(p)} \\
      \tilde{\varphi}_{1}^{(p)}
    \end{pmatrix}
    \\
    \begin{pmatrix}
      \tilde{\varphi}_{2}^{(p)} \\
      \tilde{\varphi}_{2}^{(p)}
    \end{pmatrix}\\
    \vdots \\
    \begin{pmatrix}
      \tilde{\varphi}_{N_p}^{(p)} \\
      \tilde{\varphi}_{N_p}^{(p)}
    \end{pmatrix}    
  \end{pmatrix},
  \label{holevec}
\end{eqnarray}
Eq.(\ref{Vres-mat}) can be represented as
\begin{eqnarray}
  \bvec{\mathcal{V}}
  =
  \sum_{q,q'}
  \vec{\tilde{\bvec{\varphi}}}_{q}
  \frac{\kappa_{qq'}(r)}{r^2}
  \vec{\tilde{\bvec{\varphi}}}_{q'}^{\mathsf{T}}.
  \label{Vres-mat2}
\end{eqnarray}

$\vec{\bvec{\phi}}$ is the RPA wave function which is defined by
\begin{eqnarray}
  \vec{\bvec{\phi}}
  =
  \begin{pmatrix}
    \phi_{1}^{(n)} \\
    \phi_{2}^{(n)} \\
    \vdots \\
    \phi_{N_n}^{(n)} \\
    \phi_{1}^{(p)} \\
    \phi_{2}^{(p)} \\
    \vdots \\
    \phi_{N_p}^{(p)}
  \end{pmatrix}
\end{eqnarray}
as a $2(N_n+N_p)$-dimensional vector using
\begin{eqnarray}
  \phi_{\alpha}^{(q)}
  =
  \begin{pmatrix}
    X_{\alpha}^{(q)} \\
    Y_{\alpha}^{(q)}
  \end{pmatrix}.
\end{eqnarray}

Eq.(\ref{RPAeq2}) is an $2(N_n+N_p)$-dimensional simultaneous second-order differential equation,
which can be easily solved numerically by providing appropriate boundary conditions for
the given energy using the Numerov or Runge-Kutta methods (the Numerov method is used in
this paper).
Since the $2(N_n+N_p)$-dimensional simultaneous second-order differential equations have
$2(N_n+N_p)$ types of regular and non-regular solutions, the boundary conditions for each
are given as follows.

The regular solutions $\vec{\bvec{\phi}}^{(r1;q\alpha)}$ (for $\alpha\in (1,\cdots N_q)$ and
$q=n$ and $p$) are given as the solution satisfying the boundary conditions at $r=0$
given by
\begin{eqnarray}
  \lim_{r\to 0}
  \vec{\bvec{\phi}}^{(r1;q\alpha)}
  =
  \begin{pmatrix}
    0 \\
    \vdots \\
    0 \\
    \phi_{\alpha}^{(q)}
    \to
    \begin{pmatrix}
      rj_{l_\alpha}(k_{1,\alpha}^{(q)}r) \\
      0
    \end{pmatrix}
    \\
    0 \\
    \vdots \\
    0
  \end{pmatrix}
\end{eqnarray}
The boundary conditions for the regular solutions $\vec{\bvec{\phi}}^{(r2;q\alpha)}$
(for $\alpha\in (1,\cdots N_q)$ and $q=n$ and $p$) are given by
\begin{eqnarray}
  \lim_{r\to 0}
  \vec{\bvec{\phi}}^{(r2;q\alpha)}
  =
  \begin{pmatrix}
    0 \\
    \vdots \\
    0 \\
    \phi_{\alpha}^{(q)}
    \to
    \begin{pmatrix}
      0 \\
      rj_{l_\alpha}(k_{2,\alpha}^{(q)}r) 
    \end{pmatrix}
    \\
    0 \\
    \vdots \\
    0
  \end{pmatrix}, 
\end{eqnarray}
$\vec{\bvec{\phi}}^{(r2;q\alpha)}$ corresponds to the negative energy solution
of $\vec{\bvec{\phi}}^{(r1;q\alpha)}$ because there is a relation between
$k_{1,\alpha}^{(q)}$ and $k_{2,\alpha}^{(q)}$ as $k_{2,\alpha}^{(q)}(\omega)=k_{1,\alpha}^{(q)}(-\omega)$.

The outgoing boundary conditions at the limit $r\to\infty$ for the irregular solutions
$\vec{\bvec{\phi}}^{(\pm 1;q\alpha)}$ and $\vec{\bvec{\phi}}^{(\pm 2;q\alpha)}$ are given by
\begin{eqnarray}
  \lim_{r\to \infty}
  \vec{\bvec{\phi}}^{(\pm 1;q\alpha)}
  =
  \begin{pmatrix}
    0 \\
    \vdots \\
    0 \\
    \phi_{\alpha}^{(q)}
    \to
    \begin{pmatrix}
      rh^{(\pm)}_{l_\alpha}(k_{1,\alpha}^{(q)}r) \\
      0 
    \end{pmatrix}
    \\
    0 \\
    \vdots \\
    0
  \end{pmatrix}, 
\end{eqnarray}
and 
\begin{eqnarray}
  \lim_{r\to \infty}
  \vec{\bvec{\phi}}^{(\pm 2;q\alpha)}
  =
  \begin{pmatrix}
    0 \\
    \vdots \\
    0 \\
    \phi_{\alpha}^{(q)}
    \to
    \begin{pmatrix}
      0 \\
      rh^{(\pm)}_{l_\alpha}(k_{2,\alpha}^{(q)}r) 
    \end{pmatrix}
    \\
    0 \\
    \vdots \\
    0
  \end{pmatrix}, 
\end{eqnarray}
respectively, where $h_l^{(\pm)}$ is the spherical Hankel function defined by
$h_l^{(\pm)}(kr)=j_l(kr)\pm i n_l(kr)$.

We can define the regular and irregular solution matrix $\bvec{\Phi}^{(r)}$ and $\bvec{\Phi}^{(\pm)}$ 
using the regular and irregular solutions as
\begin{eqnarray}
  \bvec{\Phi}^{(r)}
  &=&
  \begin{pmatrix}
    \vec{\bvec{\phi}}^{(r1;n1)}, &
    \vec{\bvec{\phi}}^{(r2;n1)}, &
    \cdots &
    \vec{\bvec{\phi}}^{(r1;pN_p)}, &
    \vec{\bvec{\phi}}^{(r2;pN_p)} 
  \end{pmatrix}
  \nonumber\\
  \label{regsol}
\end{eqnarray}
and
\begin{eqnarray}
  \bvec{\Phi}^{(\pm)}
  &=&
  \begin{pmatrix}
    \vec{\bvec{\phi}}^{(\pm 1;n1)}, &
    \vec{\bvec{\phi}}^{(\pm 2;n1)}, &
    \cdots &
    \vec{\bvec{\phi}}^{(\pm 1;pN_p)}, &
    \vec{\bvec{\phi}}^{(\pm 2;pN_p)} 
  \end{pmatrix}.
  \nonumber\\
  \label{iregsol}
\end{eqnarray}
These matrices are given as the $2(N_n+N_p)\times 2(N_n+N_p)$ matrix which satisfy
\begin{eqnarray}
  \left[
    \frac{\hbar^2}{2m}
    \bvec{\mathcal{K}}^2
    -    
    \left\{
    -
    \frac{\hbar^2}{2m}
    \del{^2}{r^2}
    \bvec{1}
    +
    \bvec{\mathcal{U}}
    +
    \bvec{\mathcal{V}}
    \right\}
    \right]
  \bvec{\Phi}^{(r)}
  &=&
  \bvec{0}
  \label{RPAeq-mat1}
\end{eqnarray}
and
\begin{eqnarray}
  \left[
    \frac{\hbar^2}{2m}
    \bvec{\mathcal{K}}^2
    -    
    \left\{
    -
    \frac{\hbar^2}{2m}
    \del{^2}{r^2}
    \bvec{1}
    +
    \bvec{\mathcal{U}}
    +
    \bvec{\mathcal{V}}
    \right\}
    \right]
  \bvec{\Phi}^{(\pm)}
  &=&
  \bvec{0}
  \label{RPAeq-mat2}
\end{eqnarray}
respectively. 

The Jost function $\bvec{\mathcal{J}}^{(\pm)}$ is defined as the coefficient matrix which
connects $\bvec{\Phi}^{(r)}$ and $\bvec{\Phi}^{(\pm)}$ as
\begin{eqnarray}
  \bvec{\Phi}^{(r)\mathsf{T}}
  &=&
  \frac{1}{2}
  \left[
    \bvec{\mathcal{J}}^{(+)}
    \bvec{\Phi}^{(-)\mathsf{T}}
    +
    \bvec{\mathcal{J}}^{(-)}
    \bvec{\Phi}^{(+)\mathsf{T}}
    \right].
  \label{Jostdef}
\end{eqnarray}

The Wronskian $\bvec{\mathcal{W}}^{(\pm)}$ is defined by
\begin{eqnarray}
  \bvec{\mathcal{W}}^{(\pm)}
  &=&
  \frac{\hbar^2}{2m}
  \left[
    \bvec{\Phi}^{(r)\mathsf{T}}
    \left(
    \del{}{r}
    \bvec{\Phi}^{(\pm)}
    \right)
    -
    \left(
    \del{}{r}
    \bvec{\Phi}^{(r)\mathsf{T}}
    \right)
    \bvec{\Phi}^{(\pm)}
    \right],
  \nonumber\\
  \label{wrondef}
\end{eqnarray}
and it is very easy to confirm that
this definition of Wronskian is constant for the radial coordinate $r$,
{\it i.e.} $\del{}{r}\bvec{\mathcal{W}}^{(\pm)}=0$.

By inserting Eq.(\ref{Jostdef}) into Eq.(\ref{wrondef}) and taking the
limit $r\to\infty$, it is very easy to obtain the relation formula between the Jost
function and Wronskian given by
\begin{eqnarray}
  \bvec{\mathcal{J}}^{(\pm)}
  &=&
  \pm \frac{2m}{i\hbar^2}
  \bvec{\mathcal{W}}^{(\pm)}
  \bvec{\mathcal{K}}.
  \label{jost-wron}
\end{eqnarray}
Applying the Green's theorem to Eq.(\ref{RPAeq-mat1}), we can obtain
\begin{eqnarray}
  &&
  \left[
    \bvec{\chi}^{(\pm) \mathsf{T}}(r)
    \del{}{r}
    \bvec{\Phi}^{(r)}(r)
    -
    \left(
    \del{}{r}
    \bvec{\chi}^{(\pm) \mathsf{T}}(r)
    \right)
    \bvec{\Phi}^{(r)}(r)
    \right]
  \pm i
  \bvec{\mathcal{K}}^{-1}
  \nonumber\\
  &&=
  \frac{2m}{\hbar^2}
  \int_0^r dr'
  \bvec{\chi}^{(\pm) \mathsf{T}}(r')
  \left[
    \bvec{\mathcal{U}}(r')
    +
    \bvec{\mathcal{V}}(r')
    \right]
  \bvec{\Phi}^{(r)}(r')
  \label{gr1}
\end{eqnarray}
where $\bvec{\chi}^{(\pm)}$ is the free particle wave function matrix which satisfies 
\begin{eqnarray}
  \left[
    \frac{\hbar^2}{2m}
    \bvec{\mathcal{K}}^2
    +
    \frac{\hbar^2}{2m}
    \del{^2}{r^2}
    \bvec{1}
    \right]
  \bvec{\chi}^{(\pm)}
  &=&
  \bvec{0}
  \label{freewf-hankel}
\end{eqnarray}
and the components are represented by the spherical Hankel function. 

By inserting Eq.(\ref{Jostdef}) into Eq.(\ref{gr1}) and taking the limit $r\to\infty$,
we can obtain the integral form of the Jost function
\begin{eqnarray}
  &&
  \bvec{\mathcal{J}}^{(\pm)\mathsf{T}}
  =
  \bvec{1}
  \mp
  \frac{2m}{\hbar^2}
  \frac{1}{i}
  \bvec{\mathcal{K}}
  \int_0^\infty dr'
  \bvec{\chi}^{(\pm) \mathsf{T}}(r')
  \nonumber\\
  &&\hspace{40pt}\times
  \left[
    \bvec{\mathcal{U}}(r')
    +
    \bvec{\mathcal{V}}(r')
    \right]
  \bvec{\Phi}^{(r)}(r').
\end{eqnarray}

\subsection{Perturbed Green function and RPA response function}

The Green's function is defined as a function which satisfies
\begin{eqnarray}
  &&
  \left[
    \frac{\hbar^2}{2m}
    \bvec{\mathcal{K}}^2
    -    
    \left\{
    -
    \frac{\hbar^2}{2m}
    \del{^2}{r^2}
    \bvec{1}
    +
    \bvec{\mathcal{U}}
    +
    \bvec{\mathcal{V}}
    \right\}
    \right]
  \bvec{\mathcal{G}}^{(\pm)}(r,r')
  \nonumber\\
  &&
  =
  \bvec{1}\delta(r-r').
  \label{Greendef2}
\end{eqnarray}
can be represented as
\begin{eqnarray}
  \bvec{\mathcal{G}}^{(\pm)}(r,r')
  &&=
  \theta(r-r')
  \bvec{\Phi}^{(\pm)}(r)
  \left(
  \bvec{\mathcal{W}}^{(\pm)-1}
  \right)
  \bvec{\Phi}^{(r)\mathsf{T}}(r')
  \nonumber\\
  &&\hspace{-10pt}
  +
  \theta(r'-r)
  \bvec{\Phi}^{(r)}(r)
  \left(
  \bvec{\mathcal{W}}^{(\pm)-1}
  \right)^{\mathsf{T}}
  \bvec{\Phi}^{(\pm)\mathsf{T}}(r')
  \label{Greendef1}
\end{eqnarray}
by using the Wronskian. 
Note that this Green function is also given in the form of
$2(N_n+N_p)\times 2(N_n+N_p)$ matrix 
(The proof of Eq.(\ref{Greendef1}) is shown in Appendix.\ref{Gjost}).

The RPA equation when the external field exists is given by
\begin{eqnarray}
  \left[
    \frac{\hbar^2}{2m}
    \bvec{\mathcal{K}}^2
    -    
    \left\{
    -
    \frac{\hbar^2}{2m}
    \del{^2}{r^2}
    \bvec{1}
    +
    \bvec{\mathcal{U}}
    +
    \bvec{\mathcal{V}}
    \right\}
    \right]
  \vec{\bvec{\phi}}_F
  =\vec{\bvec{F}}
  \label{RPAeq-ext}
\end{eqnarray}
where $\vec{\bvec{F}}$ is the external field which is expressed as
\begin{eqnarray}
  \vec{\bvec{F}}(r)
  &=&
  \sum_{q=n,p}
  \vec{\tilde{\bvec{\varphi}}}_{q}
  f_q(r)
  \label{extf}
\end{eqnarray}
in the form of the $2(N_n+N_p)$-dimensional vector.
The solution of the RPA equation with the external field Eq.(\ref{RPAeq-ext}) $\vec{\bvec{\phi}}_F$ 
is given by using the Green function $\bvec{\mathcal{G}}^{(\pm)}(r,r')$ as
\begin{eqnarray}
  \vec{\bvec{\phi}}_F(r)
  =
  \int_0^\infty dr'
  \bvec{\mathcal{G}}^{(+)}(r,r')
  \idot
  \vec{\bvec{F}}(r')
  \label{solphiF}
\end{eqnarray}
and the strength function is given by
\begin{eqnarray}
  S_F(\omega)
  &=&
  -\frac{1}{\pi}\mbox{ Im }
  \int dr
  \vec{\bvec{F}}^{\mathsf{T}}(r)
  \idot
  \vec{\bvec{\phi}}_F(r)
  \label{strength1}
  \\
  &=&
  -\frac{1}{\pi}\mbox{ Im }
  \int\int drdr'
  \vec{\bvec{F}}^{\mathsf{T}}(r)
  \idot
  \bvec{\mathcal{G}}^{(+)}(r,r')
  \idot
  \vec{\bvec{F}}(r')
  \nonumber\\
  \label{strength2}
\end{eqnarray}
Since the inverse of the Wronskian is included in the Green function
as shown in Eq.(\ref{Greendef1}) and the Wronskian is related to the
Jost function as shown by Eq.(\ref{jost-wron}), the pole of the strength function
may be found on the complex energy $\omega$ plane as a solution of
\begin{eqnarray}
  \det\bvec{\mathcal{J}}^{(+)}(\omega)=0.
  \label{cond0}
\end{eqnarray}

The strength function can also be expressed as
\begin{eqnarray}
  S_F(\omega)
  &=&
  -\frac{1}{\pi}
  \int dr
  \sum_{q=n,p}
  f_q(r)
  \mbox{ Im }
  \delta\rho_{F,q}(r)
  \label{strength3}
  \\
  &=&
  -\frac{1}{\pi}
  \sum_{qq'}
  \int\int drdr'
  f_q(r)
  \mbox{ Im }
  R_{qq'}(r,r')
  f_{q'}(r')
  \nonumber\\
  \label{strength4}
\end{eqnarray}
by using the density fluctuation $\delta\rho_{F,q}(r)$ and RPA response
function $R_{qq'}(r,r')$ which were defined by
\begin{eqnarray}
  \delta\rho_{F,q}(r)
  &=&
  \vec{\tilde{\bvec{\varphi}}}_{q}^{\mathsf{T}}
  \idot
  \vec{\bvec{\phi}}_F(r)
  \label{trddef}
  \\
  R_{qq'}(r,r')
  &=&
  \vec{\tilde{\bvec{\varphi}}}_{q}^{\mathsf{T}}
  \idot
  \bvec{\mathcal{G}}^{(+)}(r,r')
  \idot
  \vec{\tilde{\bvec{\varphi}}}_{q'}
  \label{resp}
\end{eqnarray}
Since $\vec{\bvec{\phi}}_F(r)$ is $2(N_n+N_p)$-dimensional vector which is represented by
\begin{eqnarray}
  \vec{\bvec{\phi}}_F
  =
  \begin{pmatrix}
    \phi_{F,1}^{(n)} \\
    \phi_{F,2}^{(n)} \\
    \vdots \\
    \phi_{F,N_n}^{(n)} \\
    \phi_{F,1}^{(p)} \\
    \phi_{F,2}^{(p)} \\
    \vdots \\
    \phi_{F,N_p}^{(p)}
  \end{pmatrix}
  \hspace{5pt}
  \mbox{ with }
  \hspace{5pt}
  \phi_{F,\alpha}^{(q)}
  =
  \begin{pmatrix}
    X_{F,\alpha}^{(q)} \\
    Y_{F,\alpha}^{(q)}
  \end{pmatrix},
\end{eqnarray}
the density fluctuation $\delta\rho_{F,q}(r)$ can be decomposed by each
transition component $\alpha$ as
\begin{eqnarray}
  \delta\rho_{F,q}(r)
  &=&
  \sum_{\alpha=1,N_q}
  \tilde{\varphi}_{\alpha}^{(q)}
  \left(X^{(q)}_{F,\alpha}(r)+Y^{(q)}_{F,\alpha}(r)\right)
  \label{trdxy}
  \\
  &=&
  \sum_{\alpha=1,N_q}
  \delta\rho_{F,q}^{(\alpha)}(r)
  \label{trda}.
\end{eqnarray}
Note that the capability of such a transition component decomposition is one of
the features of our RPA method using the Jost function. This is because the
existing cRPA method cannot perform a transition component decomposition of
density fluctuations because the density fluctuations or RPA response functions
are directly obtained.

\section{Numerical setup and check}
\label{sec3}

The first Riemann sheet on which the bound state exists is analytically connected to
the Riemann sheet on which the pole of resonance exists by a branch cut line extending
from the branching point given by the threshold energy on the real axis of complex
energy.
As will be discussed in more detail later in Sec.\ref{numrieman}, there are as many Riemann
sheets as there are sign combinations of the imaginary part of the complex momentum defining
the complex energy plane, and the number of types of complex momentum is determined by the
number of transition configurations (see Table \ref{table1} for $E1$ dipole of $^{16}$O),
so the heavier the nucleus, the more Riemann sheets that are defined. 
Therefore, in this paper the electric dipole excitations of $^{16}$O (known as relatively
light spherical nuclei) are calculated and analyzed using the Woods-Saxon potential
for the mean field $U_{lj}^{(q)}(r)$ and simple density-dependent interactions for the residual
interactions $\kappa_{qq'}$.

In this section, the model and parameters used in this paper and the results of a comparison
with the cRPA method as a numerical check will be presented in Sec.\ref{param} and
\ref{cRPA-Jost}, respectively. 
In Sec.\ref{numrieman}, an explanation of the definition of the Riemann sheet in $^{16}$O
electric dipole excitations and a numerical check of the analytic continuation is presented. 
\begin{table}
  \caption{The bound single-particle levels for neutron and proton obtained by using
  the Woods-Saxon potential model. The unit is MeV.}
  \label{table0}
  \begin{ruledtabular}
    \begin{tabular}{ccccc}
      & & Neutron & Proton &\\
      \colrule
      & $s_{1/2}$ & -36.17 & -31.16 & \\
      & $p_{3/2}$ & -21.31 & -16.84 & \\
      & $p_{1/2}$ & -16.38 & -11.95 & \\
      \\
      & $d_{5/2}$ &  -6.81 &  -2.95 & \\
      & $s_{1/2}$ &  -4.90 &  -1.43 & \\
    \end{tabular}
  \end{ruledtabular}
\end{table}

\subsection{Model and parameters}
\label{param}
The Woods-Saxon potential model and the parameters are given by
\begin{eqnarray}
  U_{lj}^{(n)}(r)
  &=&
  V_0^{(n)}f_{WS}(r)
  +
  V_1^{(n)}\bvec{l}\idot\bvec{s}\frac{1}{r}\frac{df_{WS}(r)}{dr}
  \nonumber\\
  &&
  +
  \frac{\hbar^2l(l+1)}{2mr^2}
  \\
  V_0^{(n)}
  &=&
  -60\left(1-0.67\frac{N-Z}{A}\right)
  \\
  V_1^{(n)}
  &=&
  15\left(1-0.67\frac{N-Z}{A}\right)
\end{eqnarray}
\begin{eqnarray}
  U_{lj}^{(p)}(r)
  &=&
  V_0^{(p)}f_{WS}(r)
  +
  V_1^{(p)}\bvec{l}\idot\bvec{s}\frac{1}{r}\frac{df_{WS}(r)}{dr}
  \nonumber\\
  &&
  +
  \frac{\hbar^2l(l+1)}{2mr^2}
  +
  V_C(r)
  \\
  V_0^{(p)}
  &=&
  -60\left(1+0.67\frac{N-Z}{A}\right)
  \\
  V_1^{(p)}
  &=&
  15\left(1+0.67\frac{N-Z}{A}\right)
\end{eqnarray}
\begin{eqnarray}
  f_{WS}(r)&=&\frac{1}{1+\exp((r-R)/a)}
  \\
  R&=&r_0(A-1)^{\frac{1}{3}}
  \\
  r_0&=&1.2,
  \hspace{10pt}
  a=0.65
\end{eqnarray}
\begin{eqnarray}
  V_C(r)
  =
  \left\{
  \begin{array}{cc}
    \frac{Ze^2}{2R}
    \left(3-\left(\frac{r}{R}\right)^2\right)
    & (r<R)
    \\
    \frac{Ze^2}{r} & (r\ge R)
  \end{array}
  \right.
\end{eqnarray}
We adopt and use the following residual interaction model and parameters 
\begin{eqnarray}
  \kappa_{qq}
  &=&\frac{1}{2}t_0(1-x_0)+\frac{t_3}{12}\left((5+x_3)\rho-(2+4x_3)\rho_q\right)
  \nonumber\\
  \\
  \kappa_{np}&=&\kappa_{pn}
  \nonumber\\
  &=&t_0\left(1+\frac{1}{2}x_0\right)+\frac{t_3}{12}(5+x_3)\rho
\end{eqnarray}
with $t_0=-1100$ MeV fm$^3$, $t_3=16000$ MeV fm$^6$, $x_0=0.5$ and $x_3=1.0$. 

For the external field $f_q(r)$ in Eq.(\ref{extf}), we adopt
\begin{eqnarray}
  f_n(r)&=&e\frac{Z}{A}r
  \\
  f_p(r)&=&-e\frac{N}{A}r
\end{eqnarray}
as an $E1$ dipole operator where $A=N+Z$, with $N=Z=8$ for $^{16}$O.

\begin{figure}[htbp]
\includegraphics[width=\linewidth,angle=0]{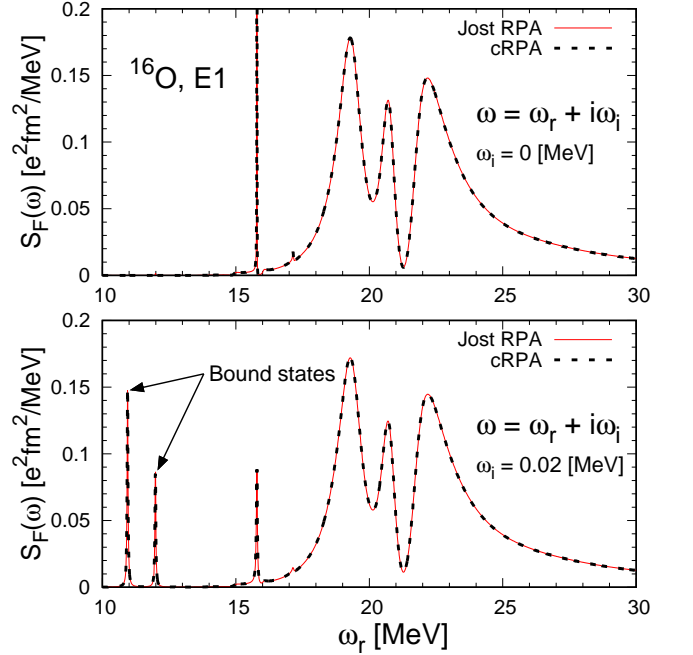}
\caption{(Color online) Comparison of the cRPA method (dotted curve) with the method
  we developed in this paper (Jost-RPA, solid curve) by the $E1$ dipole excitation strength
  of $^{16}$O.The strength function $S_F$ given as a function of complex energy
  $\omega$ is plotted as a function of the real part $\omega_r$ of complex energy.
  Note that in the upper and lower panels, the imaginary part of complex energy
  $\omega_i=0$ and $0.02$ MeV, respectively.}
\label{str_cRPA-jost}
\end{figure}

In this paper, we solved the simultaneous second-order differential equation
Eq.(\ref{RPAeq2}) by taking into account the boundary condition of continuum 
in the coordinate space up to $r\le 20$ fm with an equidistant interval
$\Delta r=0.2$ fm.

The single particle levels of neutron and proton for both
hole and particle bound states which are obtained by using the Woods-Saxon potential
are shown in Table.\ref{table0}.
The number assigned to the combination of angular momentum that couples with the angular
momentum of the hole state to make L=1 is the subscription $\alpha$ shown
in Table.\ref{table1}.
\begin{table}
  \caption{The combination of angular momenta for $L=1$ associated with the configurations of the
    particle-hole excitations in the dipole excitations of $^{16}$O and the subscript $\alpha$ that
    designates it for neutron and proton, respectively.}
  \label{table1}
  \begin{ruledtabular}
    \begin{tabular}{ccccc}
      $q$ & $\alpha$ & $[lj]^{(q)}_\alpha$ & $[l_hj_h]^{(q)}_\alpha$ & $e^{(q)}_{\alpha}$ \\
      $(=n,p)$&&&& [MeV]\\
      \colrule
      \multicolumn{5}{c}{(neutron part)}\\
       & $1$ & $s_{1/2}$ & $p_{1/2}$ & $-16.38$ \\
       & $2$ & $d_{3/2}$ & $p_{1/2}$ & $-16.38$\\
       & $3$ & $d_{5/2}$ & $p_{3/2}$ & $-21.31$ \\
     n & $4$ & $s_{1/2}$ & $p_{3/2}$ & $-21.31$\\
       & $5$ & $d_{3/2}$ & $p_{3/2}$ & $-21.31$\\
       & $6$ & $p_{3/2}$ & $s_{1/2}$ & $-36.17$\\
       & $7$ & $p_{1/2}$ & $s_{1/2}$ & $-36.17$\\
       \colrule
       \multicolumn{5}{c}{(proton part)}\\
       & $1$ & $s_{1/2}$ & $p_{1/2}$ & $-11.95$ \\
       & $2$ & $d_{3/2}$ & $p_{1/2}$ & $-11.95$\\
       & $3$ & $d_{5/2}$ & $p_{3/2}$ & $-16.84$ \\
     p & $4$ & $s_{1/2}$ & $p_{3/2}$ & $-16.84$\\
       & $5$ & $d_{3/2}$ & $p_{3/2}$ & $-16.84$\\
       & $6$ & $p_{3/2}$ & $s_{1/2}$ & $-31.16$\\
       & $7$ & $p_{1/2}$ & $s_{1/2}$ & $-31.16$\\
    \end{tabular}
  \end{ruledtabular}
\end{table}
\subsection{Comparison with cRPA method}
\label{cRPA-Jost}
The cRPA method and the Jost function method extended within the framework of the RPA theory
developed in this paper (henceforth referred to as the Jost-RPA method) are the same method
in the sense that the RPA calculation is performed by taking into account the boundary
conditions of continuum. However, the cRPA and Jost-RPA methods are completely different 
in the sense that the equations to be solved are different. 

The cRPA method calculates the unperturbed one-particle Green's function that satisfies
the boundary conditions of the continuum and then calculates the unperturbed response
function using it. The unperturbed response function is substituted into the Bethe-Salpeter
equation, which is an integral equation equivalent to the RPA equation, and the RPA response
function is obtained using the inverse matrix method in coordinate space.
In contrast, the Jost-RPA method calculates the perturbed one-particle Green's function
Eq.(\ref{Greendef1}) using the regular and non-regular solutions obtained by directly
solving the RPA equation given in the form of a differential equation Eq.(\ref{RPAeq2}),
taking into account the boundary conditions
of the continuum, and the Jost function (which is calculated using them). It is then used to
calculate the RPA response function, Eq.(\ref{resp}).

Since the two methods are thus different, we compared the two methods in the $E1$ strength
function. 
Both the cRPA and Jost-RPA methods allow the RPA response function to be calculated as a
function of continuous complex energy.
The strength function can be calculated using the RPA response function, as shown in
Eq.(\ref{strength4}).
Fig.\ref{str_cRPA-jost} shows the $E1$ strength function for $^{16}$O as a function of the
real part of the complex energy ($\omega_r$) using the cRPA and Jost-RPA methods.
In the upper and lower panels, the imaginary part of the complex energy is set to
$\omega_i=0$ and $0.02$ MeV, respectively.
The comparison of the cRPA and Jost-RPA methods in the figure shows that the results of
the cRPA and Jost-RPA methods are in perfect agreement.

Note that the peaks appearing in the energy region below threshold in the lower panel are
bound states, but these are not seen in the upper panel. This is because the poles of the
bound states appear
on the real axis of the complex energy (i.e., they have no width) and cannot be plotted as a
function of $\omega_r$ with $\omega_i=0$. 
In contrast, the strength as background is due to the contribution of the continuum, and the
resonance peaks when the width of the resonance (imaginary part of the pole) is very large
compared to $\omega_i$ are less affected by $\omega_i$ and the shape of the strength function
is almost unchanged.
These are relatively well-known properties of cRPA, and one can clearly and easily distinguish
between the bound states and the others.

Thus, the cRPA method can calculate the strength above the threshold, including the width,
but it cannot quantitatively calculate the individual width of each resonance peak.
In contrast, our Jost-RPA method can find the corresponding pole (real part is the resonance
energy, imaginary part is the resonance width) for each resonance peak individually as the zeros
of the Jost function on the complex energy Riemann sheet. The results are shown in the next
section (Sec.\ref{anal}). 

\begin{figure*}[htbp]
\includegraphics[width=0.73\linewidth,angle=-90]{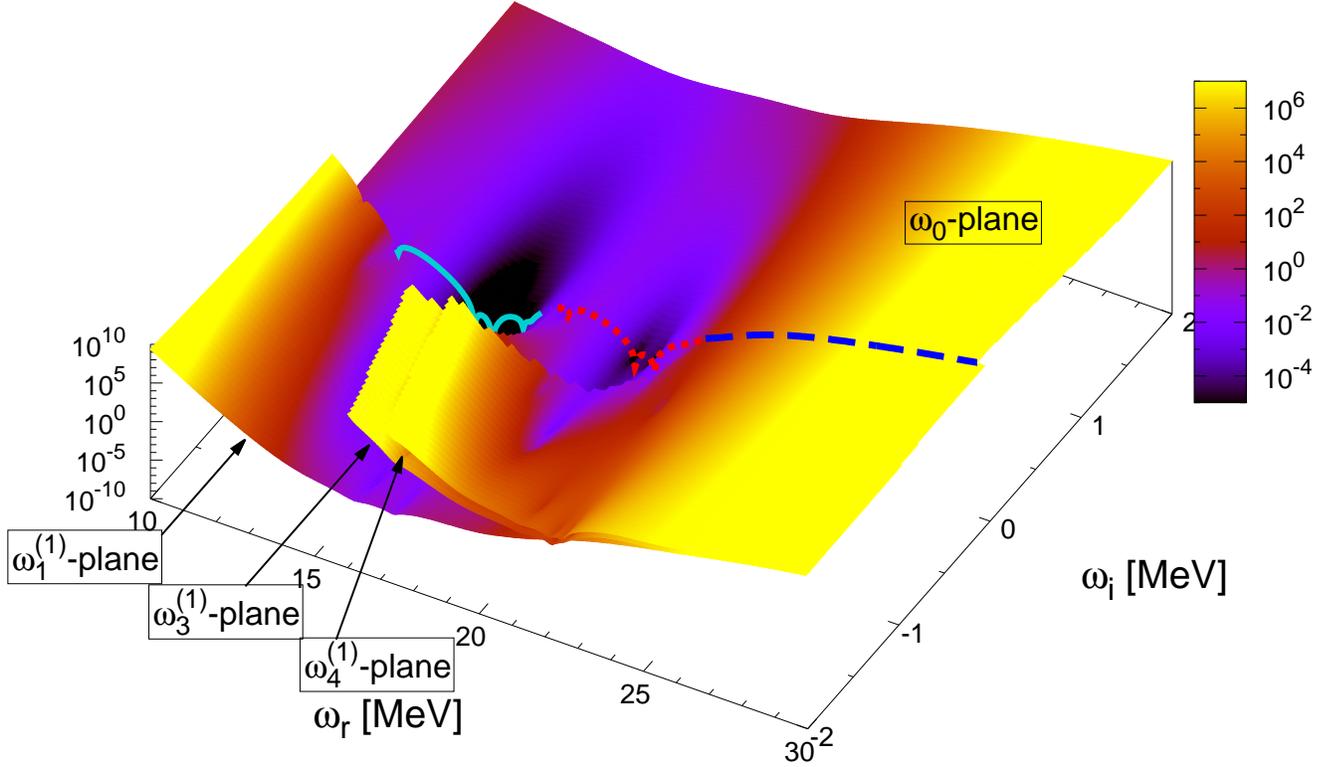}
\caption{(Color online) $|\det\bvec{\mathcal{J}}|$ plotted as a function of the
  complex energy $\omega$. The values of $|\det\bvec{\mathcal{J}}|$ with the complex
  energy defined as the first Riemann sheet ($\omega_0$) is shown in the
  Im $\omega_i > 0$ region. $|\det\bvec{\mathcal{J}}|$ with $\omega^{(1)}_1$,
  $\omega^{(1)}_3$ and $\omega^{(1)}_4$ are shown in the Im $\omega_i < 0$ region.
  And they are connected at the branch-cut lines which are existing on the real
  axis continuously. The sky-blue solid, red dotted and blue dashed curves are the branch cute
  lines which connect $\omega_0$ with $\omega^{(1)}_1$, $\omega^{(1)}_3$ and
  $\omega^{(1)}_4$, respectively.}
\label{analcont}
\end{figure*}

\subsection{Riemann sheets of the complex energy planes}
\label{numrieman}


The complex energy $\omega$ can be defined as a function of the complex momentum
$k_{1,\alpha}^{(q)}$ and $k_{2,\alpha}^{(q)}$ (Eqs.(\ref{k1def}) and (\ref{k2def})).
According to the basics of complex analysis, the types of on the complex energy plane
(Riemann sheets) are determined by the sign of the imaginary part of the complex momentum
and its combinations. 
In the case of $^{16}$O, three momentum can be defined for each of $k_{1,\alpha}^{(q)}$
and $k_{2,\alpha}^{(q)}$ for the proton and neutron, respectively, which means that 4,096
types of Riemann sheets can be defined mathematically. 
However, Riemann sheets in which the imaginary parts of $k_{1,\alpha}^{(q)}$ and $k_{2,\alpha}^{(q)}$
are simultaneously negative are unphysical. Therefore, there are a total of 128 types of physically
meaningful Riemann sheets, 64 types each corresponding to the regions Re $\omega > 0$ and
Re $\omega < 0$.
Furthermore, there are only 12 types of Riemann sheets in total, 6 each in the regions
Re $\omega > 0$ and Re $\omega < 0$, that are analytically connected on the real axis of $\omega$
with the first Riemann sheet ($\omega_0$), where the imaginary part of all momentum is defined
as positive. 
The definitions of $\omega_0$ and the 12 types of Riemann sheets are shown below.
\begin{eqnarray}
  \omega_0
  &=&
  \omega(\mbox{ Im }k_{1,\alpha}^{(q)}>0, \mbox{ Im }k_{2,\alpha}^{(q)}>0)
  \nonumber\\
  &&
  \mbox{ for all $\alpha$ and $q$}
\end{eqnarray}

\begin{eqnarray}
  \omega_1^{(s)}
  &=&
  \omega(\mbox{ Im }k_{s,\alpha}^{(q)}<0\mbox{ for }e_{\alpha}^{(q)}= -11.95)
  \nonumber\\
  &&
  \mbox{ other $\mbox{ Im }k_{1,\alpha}^{(q)}$ and $\mbox{ Im }k_{2,\alpha}^{(q)}$ are positive},
\end{eqnarray}

\begin{eqnarray}
  \omega_2^{(s)}
  &=&
  \omega(\mbox{ Im }k_{s,\alpha}^{(q)}<0\mbox{ for }e_{\alpha}^{(q)}\geq -16.38)
  \nonumber\\
  &&
  \mbox{ other $\mbox{ Im }k_{1,\alpha}^{(q)}$ and $\mbox{ Im }k_{2,\alpha}^{(q)}$ are positive},
\end{eqnarray}

\begin{eqnarray}
  \omega_3^{(s)}
  &=&
  \omega(\mbox{ Im }k_{s,\alpha}^{(q)}<0\mbox{ for }e_{\alpha}^{(q)}\geq -16.84)
  \nonumber\\
  &&
  \mbox{ other $\mbox{ Im }k_{1,\alpha}^{(q)}$ and $\mbox{ Im }k_{2,\alpha}^{(q)}$ are positive},
\end{eqnarray}

\begin{eqnarray}
  \omega_4^{(s)}
  &=&
  \omega(\mbox{ Im }k_{s,\alpha}^{(q)}<0\mbox{ for }e_{\alpha}^{(q)}\geq -21.31)
  \nonumber\\
  &&
  \mbox{ other $\mbox{ Im }k_{1,\alpha}^{(q)}$ and $\mbox{ Im }k_{2,\alpha}^{(q)}$ are positive},
\end{eqnarray}

\begin{eqnarray}
  \omega_5^{(s)}
  &=&
  \omega(\mbox{ Im }k_{s,\alpha}^{(q)}<0\mbox{ for }e_{\alpha}^{(q)}\geq -31.16)
  \nonumber\\
  &&
  \mbox{ other $\mbox{ Im }k_{1,\alpha}^{(q)}$ and $\mbox{ Im }k_{2,\alpha}^{(q)}$ are positive},
\end{eqnarray}
\begin{eqnarray}
  \omega_6^{(s)}
  &=&
  \omega(\mbox{ Im }k_{s,\alpha}^{(q)}<0\mbox{ for }e_{\alpha}^{(q)}\geq -36.17)
  \nonumber\\
  &&
  \mbox{ other $\mbox{ Im }k_{1,\alpha}^{(q)}$ and $\mbox{ Im }k_{2,\alpha}^{(q)}$ are positive},
\end{eqnarray}
where $s=1,2$. Riemann sheets $\omega_{1-6}^{(1)}$ are expected to be connected with
$\omega_0$ at $11.95 \leq \mbox{ Re }\omega \leq 16.38$ MeV,
$16.38 \leq \mbox{ Re }\omega \leq 16.84$ MeV,
$16.84 \leq \mbox{ Re }\omega \leq 21.31$ MeV,
$21.31 \leq \mbox{ Re }\omega \leq 31.16$ MeV,
and $31.16 \leq \mbox{ Re }\omega \leq 36.17$ MeV, respectively.
And $\omega_{1-6}^{(2)}$ are expected to be connected with $\omega_0$ in the
negative energy region. 
We confirmed these analytical continuation numerically in Fig.\ref{analcont}.

In Fig.\ref{analcont} we show $|\det\bvec{\mathcal{J}}|$ calculated using $\omega_0$
in the region $\omega_i > 0$, and in the region $\omega_i < 0$ we show
$|\det\bvec{\mathcal{J}}|$ calculated using the Riemann sheets 
($\omega_1^{(1)}$, $\omega_3^{(1)}$ and $\omega_4^{(1)}$-planes) connected to $\omega_0$
in the region $\omega_r < 30$ MeV
($\omega_2^{(1)}$ is also connected with $\omega_0$ in a very narrow region
$16.38  < \omega_r < 16.84$ MeV but is not shown in this figure for ease of seeing 
the figure). 

\begin{figure*}[htbp]
\includegraphics[width=0.65\linewidth,angle=-90]{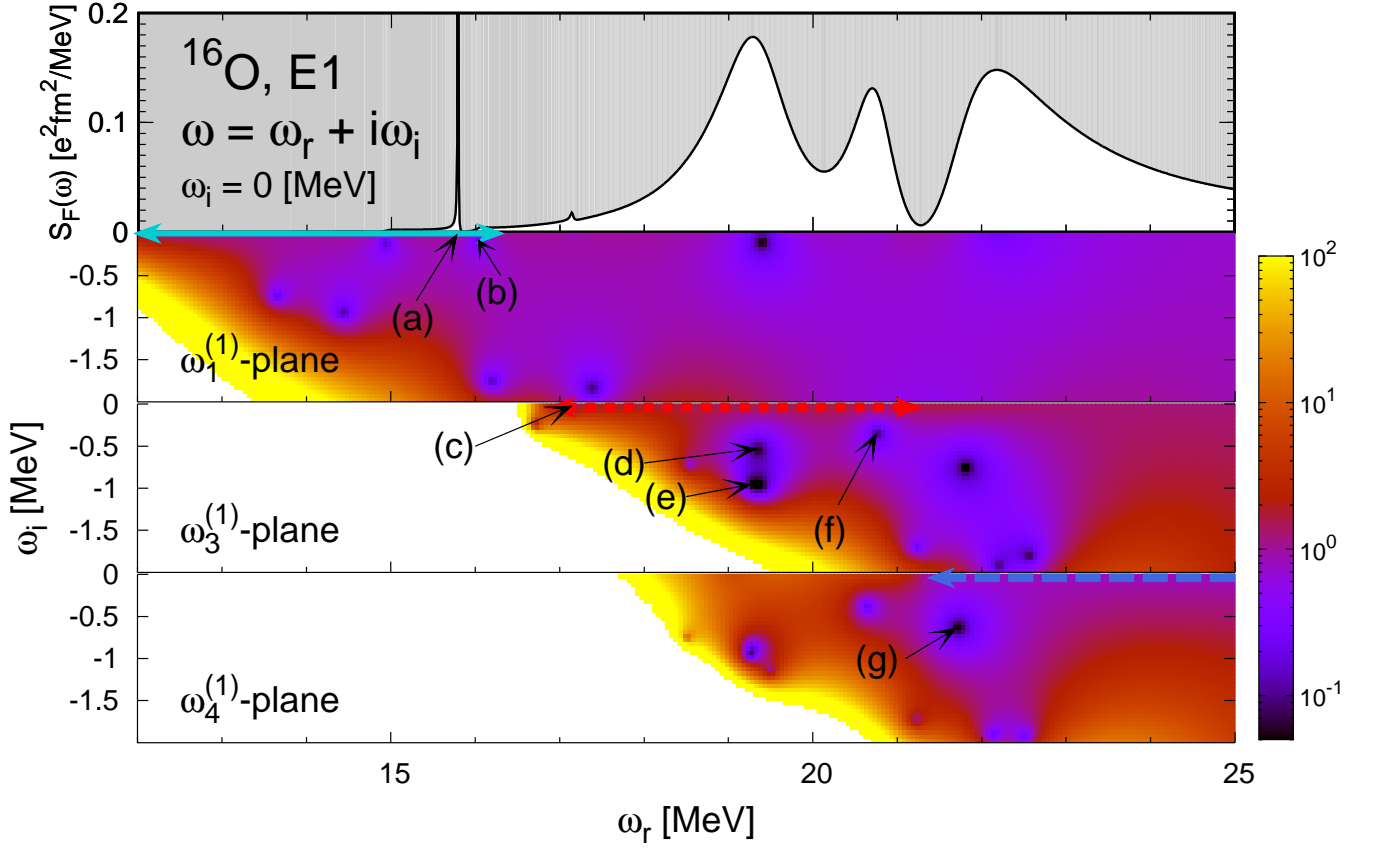}
\caption{(Color online) The $E1$ strength function of $^{16}$O plotted as a function of $\omega_r$
  with $\omega_i=0$ MeV in the upper panel, and
  $|\det\bvec{\mathcal{J}}(\omega_i^{(1)})|/|\det\bvec{\mathcal{J}}(\omega_0)|$ for $i\in 1,3$ and $4$ in
  the lower three panels. The poles labeled (a)-(g) are poles which can contribute to the strength
  function. The sky-blue solid, red dotted and blue dashed lines (with both side arrows) denote
  the branch cut lines for three Riemann sheets ($\omega_1^{(1)}$, $\omega_3^{(1)}$ and $\omega_4^{(1)}$)
  which connect with the first Riemann sheet ($\omega_0$).}
\label{pole-rpa}
\end{figure*}

\section{Analysis}
\label{anal}
\subsection{Poles of $E1$ strength of $^{16}$O on the complex energy plane}

The solution of Eq.(\ref{RPAeq2}) is given at the zero point of the Jost
function (i.e. the value of complex energy $\omega$ that satisfies Eq.(\ref{cond0}))
and the bound states appear on the real axis of complex energy below the threshold.
It is believed that the poles of the resonance state exists on the Riemann sheet that
is analytically connected to the first Riemann sheet ($\omega_0$-plane)
on the real axis above threshold. 
The real part of the resonance pole gives the resonance energy, and the imaginary part
the half-width of the resonance. 
However, the poles exhibited by $|\det\bvec{\mathcal{J}}|=0$ also include unphysical
ones due to hole-hole excited configurations, which are known to cancel at the level
of the unperturbed response function and do not contribute to the strength function,
and they appear on the real axis of the complex energy. In Fig.\ref{analcont}, such
unphysical poles appear at $\omega_r=$ 19.21 MeV and 19.79 MeV.
The presence of such unphysical poles makes it difficult to see the physical poles
that exist near the real axis when $|\det\bvec{\mathcal{J}}|$ is plotted on the
complex energy plane.
Considering the property that the poles of resonances which have widths existing
above the threshold do not appear on the real axis,
$\frac{|\det\bvec{\mathcal{J}}(\omega_i^{(1)})|}{|\det\bvec{\mathcal{J}}(\omega_0)|}$
(for $i\in 1,3,4$) is plotted instead of $|\det\bvec{\mathcal{J}}|$ in Fig.\ref{pole-rpa}
in order to find the poles corresponding to the resonances.

\begin{figure*}[htbp]
  \includegraphics[width=0.7\linewidth,angle=-90]{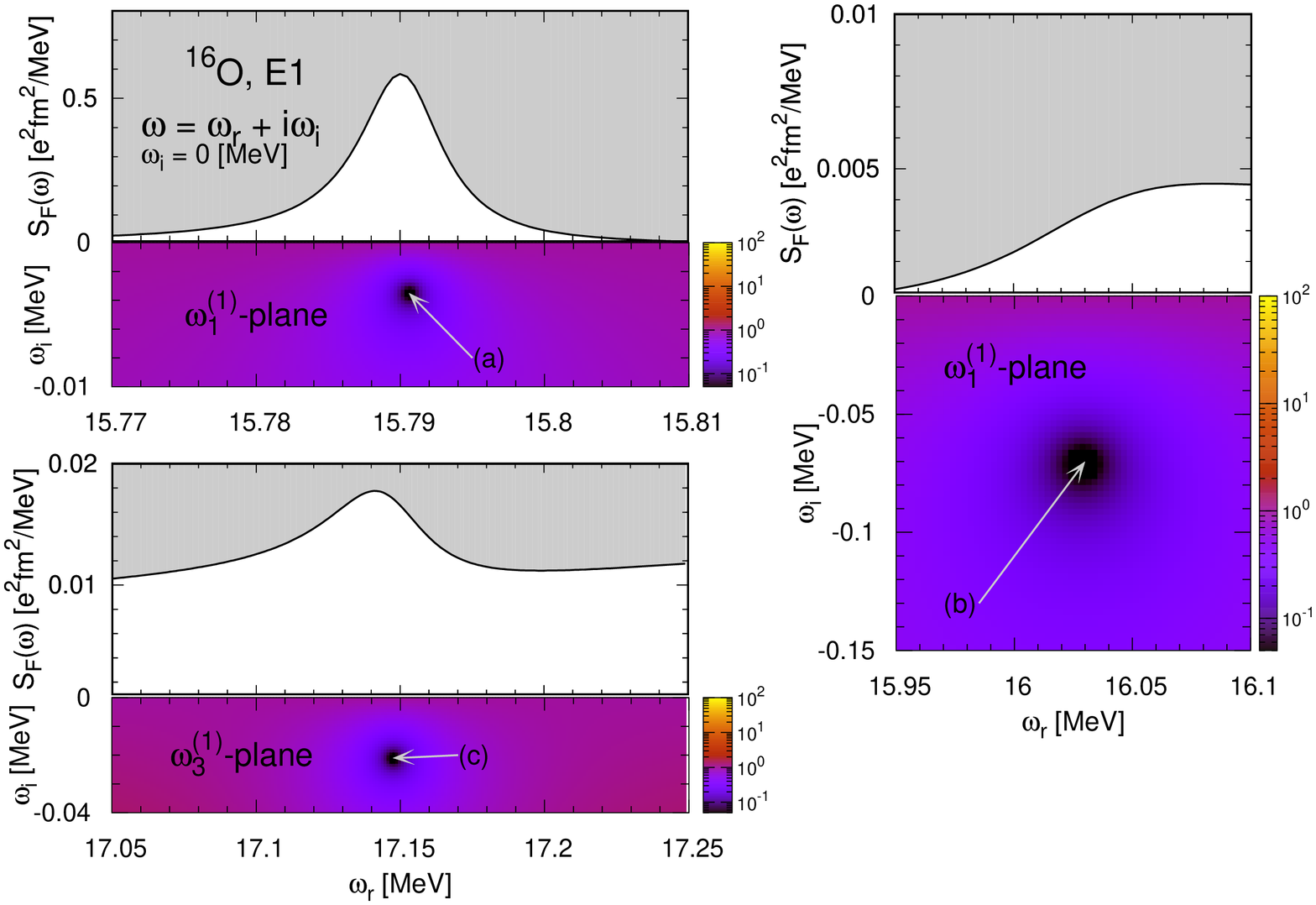}
  \caption{(Color online) In Fig. \ref{pole-rpa}, the poles (a), (b), and (c), which are located
    very close to the real axis of the complex energy plane, are shown enlarged.}
\label{pole-rpa-fine}
\end{figure*}

The top panel of Fig. \ref{pole-rpa} shows the $E1$ strength function of $^{16}$O
as a function of $\omega_r$ with $\omega_i=0$ MeV, in order to see the correspondence
between the poles on the complex energy plane and the peaks of the strength function. 
Fig.\ref{pole-rpa-fine} is an enlarged view of the region of $-0.01\leq\omega_i\leq 0.0$ MeV
and $15.77\leq\omega_r\leq 15.81$ MeV of $\omega_1^{(1)}$-plane near the low-lying peak in
Fig.\ref{pole-rpa}.
The pole that can be moved to the first Riemann sheet by rotating the branch cut line that
analytically connects the first Riemann sheet ($\omega_0$-plane) to the other Riemann sheets
on the complex energy around the branch point, e.g. by complex scaling, is the pole of
resonance, which contributes as a peak to the strength function.
The poles shown as (a)-(g) in Fig.\ref{pole-rpa} are the poles that are considered to
contribute to the peak of the strength function.
Their values of (a)-(g) are given in the second column of Table.\ref{rpasols}. 

Each of these poles found in the complex plane is considered to be linearly independent, 
and the giant electric $E1$ dipole resonance seems to be formed by four independent poles
(d), (e), (f), and (g) from Fig.\ref{pole-rpa}.
This result shows that, at least within the framework of RPA theory, one of the features
of giant resonances, the 'large width', is not given by the imaginary part of a single
pole, but is formed by several independent poles. 
In the following subsections, further analysis will be carried out to investigate the
properties of each of the poles.
\begin{figure*}[htbp]
\includegraphics[width=0.9\linewidth,angle=0]{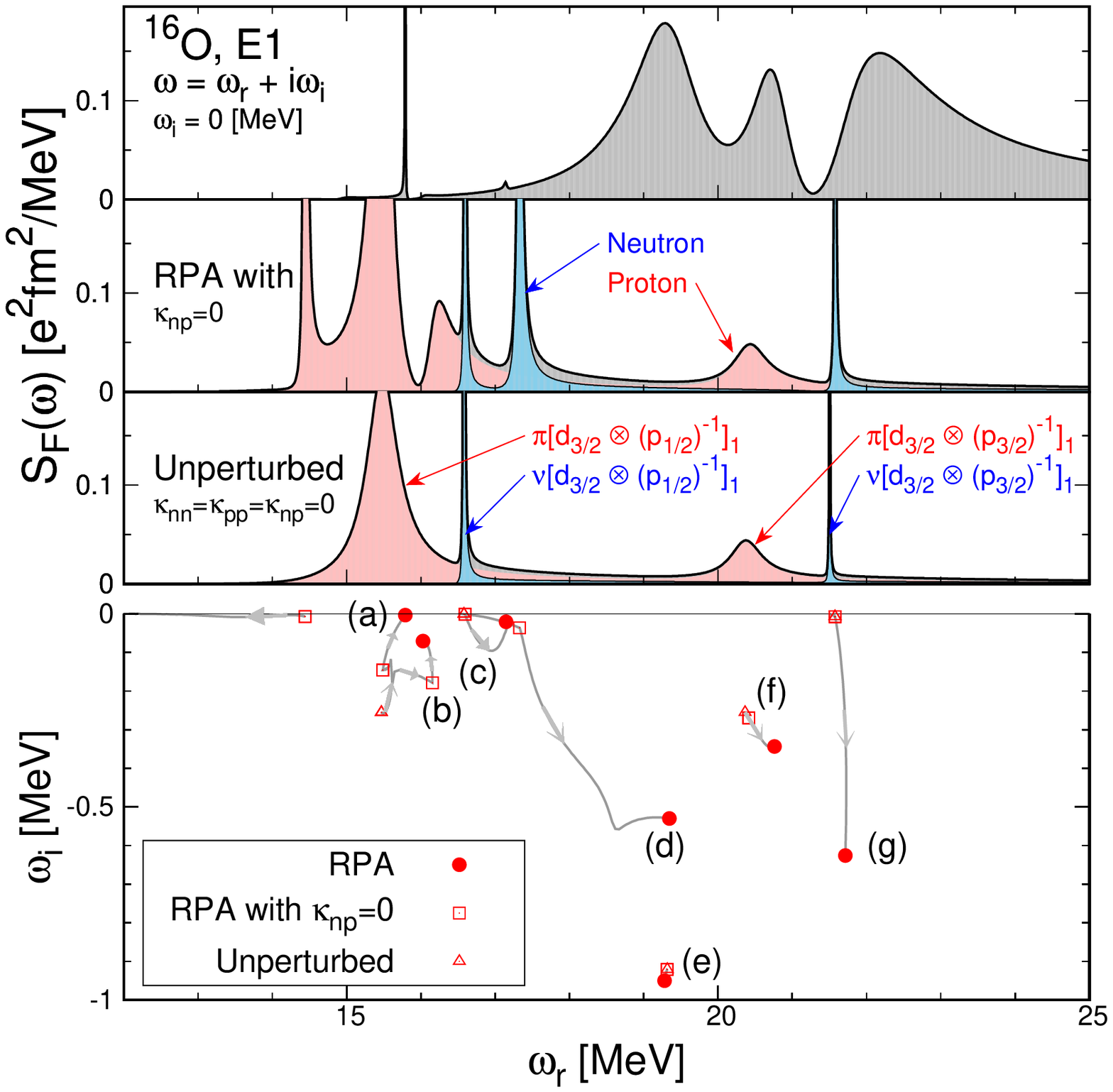}
\caption{(Color online) The $E1$ strength function which is calculated by RPA, RPA with $\kappa_{np}=0$,
  and the unperturbed response shown in the upper three panels. Trajectories of poles (a)-(g) which
  shows the connection between the RPA and unperturbed response, 
  obtained by varying the constant parameters $f_1$ and $f_2$ multiplied to the residual interaction
  are shown in the lower panel. The filled circle ($\bullet$), square ($\square$) and triangle
  ($\triangle$) symbols show the position of poles which are calculated by the RPA
  ({\it i.e. }$f_1=1,f_2=1$), RPA with $\kappa_{np}=0$ ($f_1=1, f_2=0$)
  and unperturbed ($f_1=0, f_2=0$) solutions, respectively.
  (See text and Table \ref{rpasols})}
\label{str-trj}
\end{figure*}

\begin{table*}
  \caption{The values of poles ((a)-(g)) on the complex energy planes, which are calculated
    by the RPA, RPA with $\kappa_{np}=0$ and the unperturbed response, which are corresponding to
    the filled circle ($\bullet$), square ($\square$) and triangle ($\triangle$) symbols shown in
    the lower panel of Fig.\ref{str-trj}.}
  \label{rpasols}
  \begin{ruledtabular}
    \begin{tabular}{cccccc}
       Pole & RPA & RPA with $\kappa_{np}=0$ & Unperturbed($\kappa_{qq'}=0$) &  Riemann surface &\\
       & [MeV] & [MeV] & [MeV] &  &\\
      \colrule
      (a) & $15.79-i 3.35\times 10^{-3}$ & $15.48-i 0.15$ & $15.47-i 0.26$& $\omega_1^{(1)}$-plane & \\
      (b) & $16.03-i 0.07$ & $16.16-i 0.18$ & $15.47-i0.26$ & \\
      \colrule
      (c) & $17.15-i 0.02$ & $16.59-i 1.39\times 10^{-3}$ & $16.58-i 1.33\times 10^{-3}$ & $\omega_3^{(1)}$-plane \\
      (d) & $19.34-i 0.53$ & $17.32-i 3.64\times 10^{-2}$ & $16.58-i 1.33\times 10^{-3}$ & \\
      (e) & $19.28-i 0.95$ & $19.32-i 0.92$ & $19.32-i 0.92$ & \\
      (f) & $20.76-i 0.34$ & $20.41-i 0.27$ & $20.36-i 0.26$ & \\
      \colrule
      (g) & $21.71-i 0.63$ & $21.58-i 7.45\times 10^{-3}$ & $21.50-i 1.33\times 10^{-3}$ & $\omega_4^{(1)}$-plane\\
    \end{tabular}
  \end{ruledtabular}
\end{table*}

\subsection{Trajectories of poles as a response to residual interactions}

Multiplying the residual interactions $\kappa_{qq}$ (for both $\kappa_{nn}$ and $\kappa_{pp}$) and
$\kappa_{np}$ by constants $f_1$ and $f_2$, respectively, as $\kappa_{qq}\to f_1\times \kappa_{qq}$,
$\kappa_{np}\to f_2\times\kappa_{np}$, and varying them from 0 to 1, the trajectories of the
poles can be drawn as shown in the bottom panel of Fig.\ref{str-trj}.
The top three panels show, in order from top to bottom, the RPA strength function ($f_1=f_2=1$),
the RPA strength function with $f_1=1$ and $f_2=0$ (i.e. $\kappa_{np}=0$) and the unperturbed strength
function ($f_1=f_2=0$, i.e. $\kappa_{qq'}=0$).
The positions of the poles corresponding to the peaks of the RPA strength function,
the RPA strength function with $\kappa_{np}=0$ and the unperturbed strength function
are shown by circle ($\bullet$), square ($\square$) and triangle ($\triangle$) symbols
respectively in the bottom panel.
The specific values of the circle ($\bullet$), square ($\square$) and triangle ($\triangle$)
for the poles in (a)-(g) are given in columns 2, 3 and 4 of Table \ref{rpasols}, respectively.

Focus on poles (d), (e), (f) and (g), which seem to be related to giant resonances.
It can be seen from Fig.\ref{str-trj} that, first of all, poles (d), (e), (f) and (g) have
different origins.
The pole (d) arises from the pole of the unperturbed neutron resonance of
$\nu\left[d_{3/2}\otimes\left(p_{1/2}\right)^{-1}\right]_1$ at $16.58-i1.33\times 10^{-3}$ MeV. 
This unperturbed neutron resonance appears as a very sharp peak in the unperturbed strength
function.
The pole (e) is almost unaffected by residual interactions and no corresponding peak in the
unperturbed strength function. This is thought to be a shape resonance created by a mean
field with a very wide width and only a small contribution as background of the strength
function.
The pole (f) originates from the unperturbed proton resonance of
$\pi\left[d_{3/2}\otimes\left(p_{3/2}\right)^{-1}\right]_1$ at $20.36-i0.26$ MeV 
and appears in the unperturbed strength function as a peak with a rather broad width. 
The effect of residual interactions is small. 
The pole (g) arises from the pole of the unperturbed neutron resonance of
$\nu\left[d_{3/2}\otimes\left(p_{3/2}\right)^{-1}\right]_1$ at $21.50-i1.3\times 10^{-3}$ MeV. 
This unperturbed neutron resonance appears as a very sharp peak in the unperturbed strength
function.

The influence of the residual interactions is remarkable for poles (d) and (g), both of
which originate from the poles of the neutron unperturbed resonance.
As $\kappa_{np}$ increases, both the resonance energy (real part of the pole) and the width
(imaginary part of the pole) of the pole (d) increase significantly.
The significant shift of the resonance energy of the $E1$ dipole resonance to higher energy
due to residual interactions is a typical property of the collective excited state shown
by the schematic model~\cite{ring}.
However, the response property of the pole (d) to the residual interaction is not only
the resonance energy (the real part of the pole) shift to higher energy, but also the
significant increase in the width (the imaginary part of the pole). 
In the case of the pole (g), the response to the residual interaction shows almost no energy
shift, but a remarkable increase in the width.
This implies that this is mainly due to the dominance of coupling with the non-resonant continuum.
And this may be related to the fact that only the pole (g) lies on a different Riemann sheet
from the other resonances, and there is no proton-originated pole on
the same Riemann sheet (see Fig.\ref{pole-rpa}). 

Poles (d) and (f) originate from the poles of the neutron and proton unperturbed resonances,
respectively, so they originally belong to different Riemann sheets. However, the $\kappa_{np}$
effect mixes the neutron and proton components so that (d) and (f) are finally poles belonging to
the same Riemann sheet. However, pole (f) has a very small energy shift, and its width does
not change much compared to pole (d). This may be due to the fact that the unperturbed
resonance (shape resonance) character tends to remain stronger at pole (f) due to the Coulomb
barrier.

\subsection{Component structure analysis of density fluctuations}

\begin{figure}[htbp]
\includegraphics[width=\linewidth,angle=0]{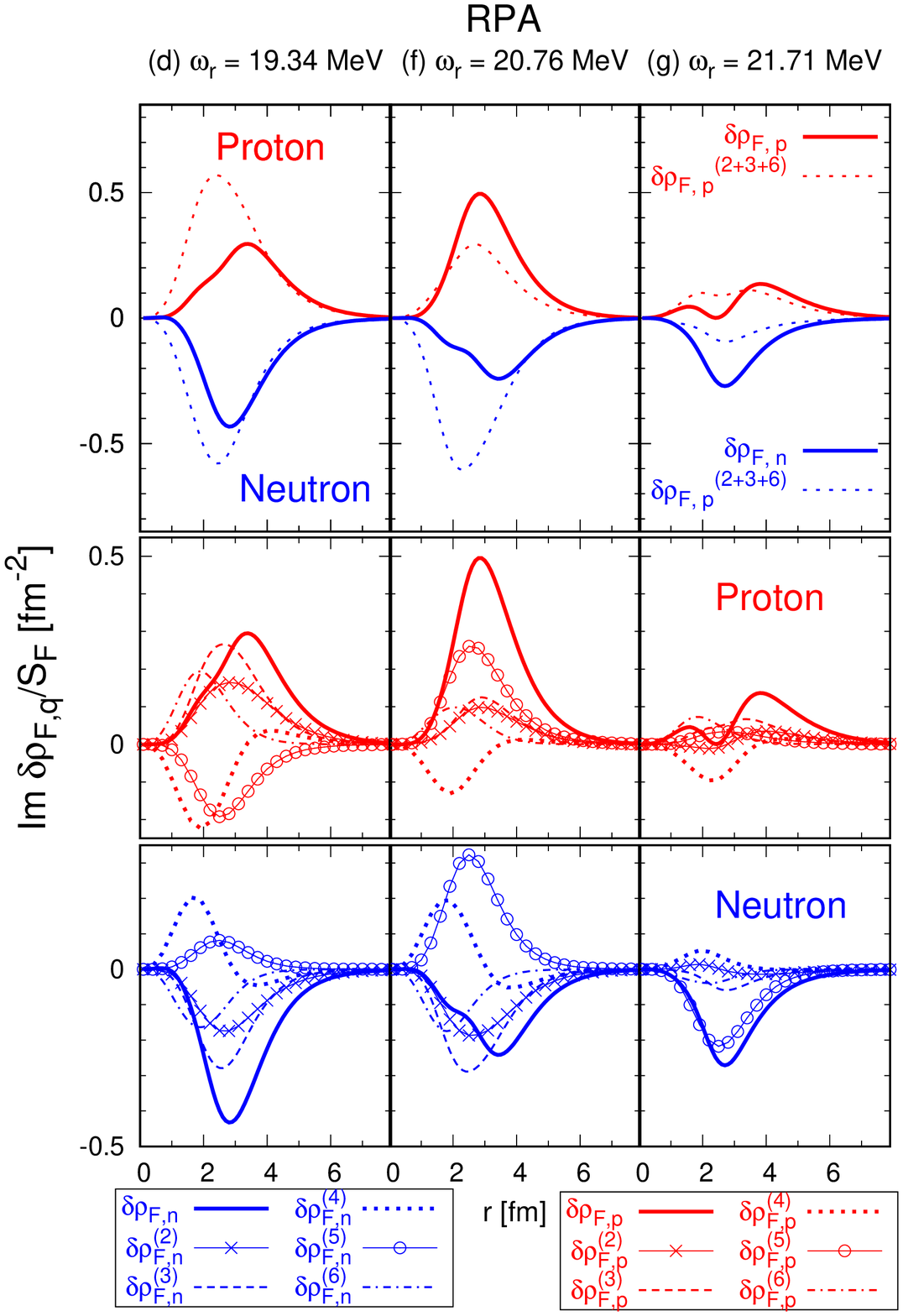}
\caption{(Color online) The density fluctuations $\delta\rho_F$ defined by Eq.(\ref{trddef})
  for the poles (d), (f) and (g) which are expected to be related with the giant dipole
  resonance are shown in the top panels by the red solid (for proton) and blue solid
  (for neutron) curves. The dotted curves show the summation of the $\alpha=2$, $3$ and $6$
  components. Component decomposition of the density fluctuation for proton and neutron
  are shown in the middle and bottom panels, respectively. }
\label{trd-RPA}
\end{figure}

\begin{figure}[htbp]
\includegraphics[width=\linewidth,angle=0]{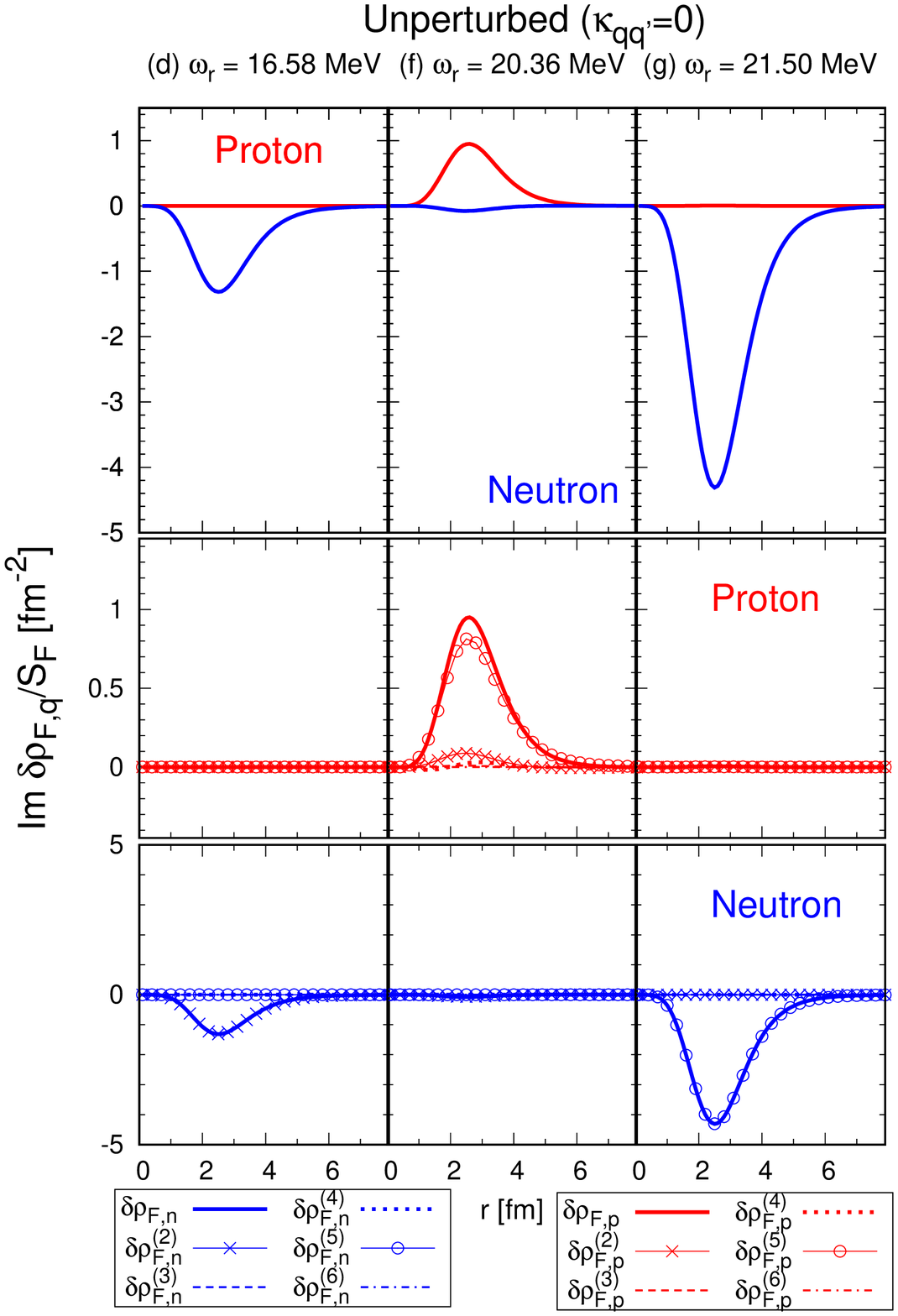}
\caption{(Color online) The same figure with Fig.\ref{trd-RPA} but calculated with the
  unperturbed response ($\kappa_{qq'}=0$).}
\label{trd-Unp}
\end{figure}

In the Jost-RPA method, the density fluctuation $\delta\rho_{F,q}(r)$ is
defined by Eq.(\ref{trddef}) and can be decomposed into components with
subscription $\alpha$, as shown in Eq.(\ref{trda}).
Fig. \ref{trd-RPA} and \ref{trd-Unp} show the imaginary part of
the RPA density fluctuations and unperturbed density fluctuations for poles
(d), (f) and (g), respectively.
The energy of the density fluctuations is the real part of the poles,
{\it i.e.}, on the real axis of the complex energy plane ($\omega_i=0$).
In each figure, the density fluctuations for protons and neutrons are shown
in the upper panel by solid curves, while the density fluctuations for
the $\alpha=2,3$ and $6$ components added together are shown by dotted
curves.
In the middle and lower panels, the main contributing components are shown when
the proton and neutron density fluctuations are decomposed into transition
components, respectively. 
Note that the density fluctuations shown in these figures are normalized by the
value of the strength function $S_F$ at a given energy ($\omega_r$).

Looking at the solid curves in the top panels of Fig. \ref{trd-RPA}, the density
fluctuations at poles (d) and (f) both roughly show the neutron density and proton
density oscillating in antiphase, which is a typical feature of the collective
motion of isovector dipole excitations.
The density fluctuations of the pole (g) also show movement of the neutron and
proton densities in antiphase, but the proton density is likely to jump out of
the nucleus.

In the case of the pole (d) (originating from the neutron unperturbed transition
$\alpha=2$, i.e. $\nu\left[d_{3/2}\otimes\left(p_{1/2}\right)^{-1}\right]_1$),
both proton and neutron components ($\delta\rho_{F,p}$ and $\delta\rho_{F,n}$) consist
of a superposition of in-phase $\alpha=2, 3$ and $6$
components, plus $\alpha=4$ and $5$ components with antiphase.
In the case of the pole (f) (originating from the proton unperturbed transition
$\alpha=5$, i.e. $\pi\left[d_{3/2}\otimes\left(p_{3/2}\right)^{-1}\right]_1$),
the $\alpha=5$ component in the proton component $\delta\rho_{F,p}$ is
enhanced by the in-phase superposition of the $\alpha=2, 3$ and $6$ components,
to which the antiphase $\alpha=4$ component is added.
The neutron components $\delta\rho_{F,n}$) consists of a superposition of in-phase
$\alpha=2, 3$ and $6$ components, plus $\alpha=4$ and $5$ components with
antiphase.
In the case of the pole (g) (originating from the neutron unperturbed transition
$\alpha=5$, i.e. $\nu\left[d_{3/2}\otimes\left(p_{3/2}\right)^{-1}\right]_1$),
the proton component $\delta\rho_{F,p}$ is in the form that the intranuclear part
($r<3$ fm region) of the amplitude created by the in-phase superposition of the
$\alpha=2, 3$ and $6$ components is damped by the $\alpha=4$ component, leaving
mainly only the extranuclear component.
The neutron component $\delta\rho_{F,n}$ is dominated by the $\alpha=5$
component and the influence of the other components is relatively small. 

The component structure of the density fluctuations of pole (d) is roughly antiphase
to each other for neutrons and protons, whereas the density fluctuations of pole (f)
have a neutron component very similar to that of pole (d), but the proton component
does not have the antiphase structure of neutrons. This may be because the proton
component still keeps its unperturbed resonance (shape resonance) character
due to the Coulomb barrier, although poles (d) and (f) influence each other by
belonging to the same Riemann sheet.
Or it could be that pole (d) and pole (f) overlap each other due to their wide widths, 
and the component structure of pole (d) just appears in the component structure of
the density fluctuations of pole (f), but actually pole (f) itself is a shape
resonance with little effect from residual interactions.

The component structure of the density fluctuations of the pole (g) is very different
from that of the poles (d) and (f). This may be due to the fact that pole (g) belongs
to a different Riemann sheet than (d) and (f). The difference in component structure
between protons and neutrons is assumed to be due to the fact that only unperturbed
resonances (shape resonances) of neutrons exist in the same Riemann sheet.
This may mean that, when the $E1$ excitation occurs, in which the neutron and proton
move in opposite phases, the neutron tends to stay in the nucleus due to the
characteristic of shape resonance, while the proton mainly couples to the nonresonant
continuum and tends to move away from the nucleus.

These are inferences based on the analysis which we have done in this paper. In order to
clarify whether the inferences are correct, it is necessary to analyze the pole-by-pole
contribution to the strength function and density fluctuations, excluding overlaps from
other resonances, etc. However, this requires changing the completeness of the system
using the complex scaling method~\cite{aoyama} or Berggren's method~\cite{berggren},
because the completeness of the system is defined by the first Riemann
sheet~\cite{newton} (The contribution of resonances is not explicitly included in the
completeness). These issues are for the future.

For the moment, the analysis at the current stage of this paper has shown that poles
(d), (f) and (g), which may be related to the $E1$ giant dipole resonance, basically
differ in their properties (resonance energy, width, response to residual interactions,
component structure of density fluctuations) for each pole. Especially when a pole
belongs to a different Riemann sheet from the others (e.g. pole (g)), the properties
are more markedly different.

\section{Summary and perspective}
\label{sec4}

In this paper, the Jost function method is extended within the framework of RPA theory
to find the complex eigenvalues of the RPA equation on the complex energy plane
(Jost-RPA method).
As a first application of the Jost-RPA method, we chose $^{16}$O electric dipole excitations
and performed numerical calculations using the Woods-Saxon potential as the mean field
and simple density-dependent interactions as the residual interactions. 
The cRPA and Jost-RPA methods are identical in terms of solving the RPA problem by considering
the boundary conditions of the continuum, but the methods for solving the equations are
completely different. Therefore, we first compared both methods by calculating the $E1$ strength
function using the same potentials and residual interactions. It was found that the result of
the Jost-RPA method is in perfect agreement with the result of the cRPA method.

In the original Jost function method, it is known that the complex energy eigenvalues of the
system are obtained as the zeros of the Jost function on the complex energy plane that is
analytically connected to the first Riemann sheet on the branch cut line.
In the Jost-RPA method, a very large number of complex-energy Riemann sheets are defined,
even for $^{16}$O $E1$ excitations. However, only a very limited number of Riemann sheets are
analytically connected to the first Riemann sheet by a branch cut line with a branch point
at the threshold of each configuration on the real axis of complex energy.
We checked numerically whether the Jost functions computed on those Riemann sheets actually
have an analytical connection with the first Riemann sheet. 
Then, the poles (complex energy eigenvalues) corresponding to the peaks of the $E1$ strength
function were successfully found numerically on the complex energy plane in analytical
connection with the first Riemann sheet.

The poles found on the complex energy plane revealed that the $E1$ giant dipole
resonance of $^{16}$O is formed by three independent poles with different resonance energies
and widths. Two of these three poles belong to the same Riemann sheet, and only one belongs
to a different Riemann sheet.
Trajectory analysis of the poles in terms of their response to residual interactions also shows
that these three poles originate from different poles of unperturbed resonance and that the
character of their response to residual interactions is also different for each pole. 
It was found that response characteristics to the residual interaction differed from pole to
pole: a pole shifted towards higher energies with increasing width as the residual interaction
became stronger, another pole was less affected by the residual interaction, and another pole
only increased in width.
The component structure of the density fluctuations is also characteristic for each pole, but
the interpretation is not clear because of the possibility of overlap due to the width of the
poles (especially for poles belonging to the same Riemann sheet). Only one thing is clear:
the poles belonging to different Riemann sheets seem to have little influence on each other
and to have quite different properties.

Based on the residue theorem in complex function theory, it is possible to extract and show
the contribution of a specific pole in the complex plane in terms of physical quantities such
as density fluctuations or strength functions. However, the completeness of the system is defined
by the first Riemann sheet, and the resonance poles are not explicitly included in the completeness.
Therefore, in order to apply  the residue theorem for more detailed analysis, the completeness
must be modified using the complex scaling method or Berggren's method. Further analysis by
improving the Jost-RPA method in this direction is expected to clarify points that were unclear
in the analysis of this paper. 

In this paper, RPA calculations for $E1$ dipole excitations of $^{16}$O have been performed
using the 
Woods-Saxon potential for the mean field and simple density-dependent interactions for the residual
interactions, but other excitation modes and other nuclei are naturally of interest.
However, for a more quantitatively reliable analysis, it is also important to perform
self-consistent calculations~\cite{sagawa,terasaki,colo} based on effective two-body nuclear
forces, such as the Skyrme interactions~\cite{SkII,SIII,SGII,SkMs,SkP,SLy4,SkO}.
This is because it is well known that calculations that break the self-consistency 
do not satisfy the sum rule for dipole excitations~\cite{sil},
and there is no phenomenological potential model for cases such as neutron-rich nuclei.
Removing spurious modes is another important issue~\cite{ring,ripka,lane,rpa-fam}. 
In this paper, an approximation to remove
the spurious mode from the external field was used, but this approximation may not work well
in some cases, so it is preferable to remove it directly from the RPA solution.
The cRPA has already developed a method to remove the spurious mode from the
RPA response function~\cite{sup-crpa}.
A method to remove the spurious mode directly from the solution should also be developed
for the Jost-RPA method.

We are planning to continue our research using the Jost-RPA method developed in this paper, improving 
the above-mentioned issues step by step, in order to further analyze and understand the excited
structure of nuclei in more detail from the viewpoint of the resonance poles in the near future.

\section*{Acknowledgments}
This work is funded by Vietnam National Foundation for Science and Technology
Development (NAFOSTED) under grant number “103.04-2019.329”. 
Tran Dieu Thuy was funded by the Master, PhD Scholarship Programme of Vingroup Innovation
Foundation (VINIF), code VINIF.2022.TS128. This work was partially supported by the Hue University under the Core Research Program, grant No. NCM.DHH.2018.09.

\appendix

\subsection{Proof of Eq.(\ref{Greendef1})}
\label{Gjost}

The Green's theorem can lead the following equations
\begin{eqnarray}
  &&
  \left[
    \bvec{\Phi}_0^{(r) \mathsf{T}}(r)
    \del{}{r}
    \bvec{\Phi}^{(r)}(r)
    -
    \left(
    \del{}{r}
    \bvec{\Phi}_0^{(r) \mathsf{T}}(r)
    \right)
    \bvec{\Phi}^{(r)}(r)
    \right]
  \nonumber\\
  &&=
  \frac{2m}{\hbar^2}
  \int_0^r dr'
  \bvec{\Phi}_0^{(r) \mathsf{T}}(r')
  \bvec{\mathcal{V}}(r')
  \bvec{\Phi}^{(r)}(r')
  \label{gr01}
\end{eqnarray}
\begin{eqnarray}
  &&
  \pm i
  \bvec{\mathcal{K}}^{-1}
  \bvec{\mathcal{J}}^{(\pm)}_0
  \nonumber\\
  &&
  -
  \left[
    \bvec{\Phi}_0^{(r) \mathsf{T}}(r)
    \del{}{r}
    \bvec{\Phi}^{(\pm)}(r)
    -
    \left(
    \del{}{r}
    \bvec{\Phi}_0^{(r) \mathsf{T}}(r)
    \right)
    \bvec{\Phi}^{(\pm)}(r)
    \right]
  \nonumber\\
  &&=
  \frac{2m}{\hbar^2}
  \int_r^\infty dr'
  \bvec{\Phi}_0^{(r) \mathsf{T}}(r')
  \bvec{\mathcal{V}}(r')
  \bvec{\Phi}^{(\pm)}(r')
  \label{gr03}
\end{eqnarray}
where $\bvec{\Phi}_0^{(r)}(r)$ and $\bvec{\Phi}_0^{(\pm)}(r)$ are
regular and irregular solution matrix of 
\begin{eqnarray}
  \left[
    \frac{\hbar^2}{2m}
    \bvec{\mathcal{K}}^2
    -    
    \left\{
    -
    \frac{\hbar^2}{2m}
    \del{^2}{r^2}
    \bvec{1}
    +
    \bvec{\mathcal{U}}
    \right\}
    \right]
  \bvec{\Phi}_0^{(r)}
  &=&
  \bvec{0}
  \label{HFeq-mat1}
\end{eqnarray}
and
\begin{eqnarray}
  \left[
    \frac{\hbar^2}{2m}
    \bvec{\mathcal{K}}^2
    -    
    \left\{
    -
    \frac{\hbar^2}{2m}
    \del{^2}{r^2}
    \bvec{1}
    +
    \bvec{\mathcal{U}}
    \right\}
    \right]
  \bvec{\Phi}_0^{(\pm)}
  &=&
  \bvec{0},
  \label{HFeq-mat2}
\end{eqnarray}
respectively. And $\bvec{\mathcal{J}}^{(\pm)}_0$ is the Jost function
which is defined by $\bvec{\Phi}_0^{(r)}(r)$ and $\bvec{\Phi}_0^{(\pm)}(r)$. 
It should be noted that
$\bvec{\Phi}_0^{(r)}(r)$, $\bvec{\Phi}_0^{(\pm)}(r)$ and $\bvec{\mathcal{J}}^{(\pm)}_0$
are given as the $2(N_n+N_p)\times 2(N_n+N_p)$ diagonal matrix due to the absence
of the residual interaction $\bvec{\mathcal{V}}$ which has the
off-diagonal components.

Introducing a $2(N_n+N_p)\times 2(N_n+N_p)$ dimensional matrix $\bvec{\mathcal{C}}$
and calculating $\bvec{\Phi}^{(\pm)}(r)\bvec{\mathcal{C}}\times$(Eq.(\ref{gr01}))$^{\mathsf{T}}$
+$\bvec{\Phi}^{(r)}(r)\bvec{\mathcal{C}}^{\mathsf{T}}\times$(Eq.(\ref{gr03}))$^{\mathsf{T}}$, 
we can obtain
\begin{eqnarray}
  &&
  \pm i
  \bvec{\Phi}^{(r)}(r)
  \bvec{\mathcal{C}}^{\mathsf{T}}
  \bvec{\mathcal{J}}^{(\pm)}_0
  \bvec{\mathcal{K}}^{-1}
  \nonumber\\
  &&
  +
  \left[
    \bvec{\Phi}^{(\pm)}(r)
    \bvec{\mathcal{C}}
    \left(
    \del{}{r}
    \bvec{\Phi}^{(r)\mathsf{T}}(r)
    \right)
    \right.
    \nonumber\\
    &&\hspace{40pt}
    \left.
    -
    \bvec{\Phi}^{(r)}(r)
    \bvec{\mathcal{C}}^{\mathsf{T}}
    \left(
    \del{}{r}
    \bvec{\Phi}^{(\pm)\mathsf{T}}(r)
    \right)
    \right]
  \bvec{\Phi}_0^{(r)}(r)
  \nonumber\\
  &&
  -
  \left[
    \bvec{\Phi}^{(\pm)}(r)
    \bvec{\mathcal{C}}
    \bvec{\Phi}^{(r)\mathsf{T}}(r)
    \right.
    \nonumber\\
    &&\hspace{40pt}
    \left.
    -
    \bvec{\Phi}^{(r)}(r)
    \bvec{\mathcal{C}}^{\mathsf{T}}
    \bvec{\Phi}^{(\pm)\mathsf{T}}(r)
    \right]
    \left(
    \del{}{r}
    \bvec{\Phi}_0^{(r)}(r)
    \right)
  \nonumber\\
  &&=
  \frac{2m}{\hbar^2}
  \bvec{\Phi}^{(r)}(r)
  \bvec{\mathcal{C}}^{\mathsf{T}}
  \int_r^\infty dr'
  \bvec{\Phi}^{(\pm)\mathsf{T}}(r')
  \bvec{\mathcal{V}}(r')
  \bvec{\Phi}_0^{(r)}(r')
  \nonumber\\
  &&
  +
  \frac{2m}{\hbar^2}
  \bvec{\Phi}^{(\pm)}(r)
  \bvec{\mathcal{C}}
  \int_0^r dr'
  \bvec{\Phi}^{(r)\mathsf{T}}(r')
  \bvec{\mathcal{V}}(r')
  \bvec{\Phi}_0^{(r)}(r')
  \label{gr013}
\end{eqnarray}
If we require
\begin{eqnarray}
  &&
  \left[
    \bvec{\Phi}^{(\pm)}(r)
    \bvec{\mathcal{C}}
    \left(
    \del{}{r}
    \bvec{\Phi}^{(r)\mathsf{T}}(r)
    \right)
    \right.
    \nonumber\\
    &&\hspace{40pt}
    \left.
    -
    \bvec{\Phi}^{(r)}(r)
    \bvec{\mathcal{C}}^{\mathsf{T}}
    \left(
    \del{}{r}
    \bvec{\Phi}^{(\pm)\mathsf{T}}(r)
    \right)
    \right]
  =-\bvec{1}
  \label{gr013-c1}
  \\
  &&
  \left[
    \bvec{\Phi}^{(\pm)}(r)
    \bvec{\mathcal{C}}
    \bvec{\Phi}^{(r)\mathsf{T}}(r)
    \right.
    \nonumber\\
    &&\hspace{40pt}
    \left.
    -
    \bvec{\Phi}^{(r)}(r)
    \bvec{\mathcal{C}}^{\mathsf{T}}
    \bvec{\Phi}^{(\pm)\mathsf{T}}(r)
    \right]
  =\bvec{0},
  \label{gr013-c2}
\end{eqnarray}
then we can rewrite Eq.(\ref{gr013}) as
\begin{eqnarray}
  &&
  \pm i
  \bvec{\Phi}^{(r)}(r)
  \bvec{\mathcal{C}}^{\mathsf{T}}
  \bvec{\mathcal{J}}^{(\pm)}_0
  \bvec{\mathcal{K}}^{-1}
  -
  \bvec{\Phi}_0^{(r)}(r)
  \nonumber\\
  &&=
  \frac{2m}{\hbar^2}
  \bvec{\Phi}^{(r)}(r)
  \bvec{\mathcal{C}}^{\mathsf{T}}
  \int_r^\infty dr'
  \bvec{\Phi}^{(\pm)\mathsf{T}}(r')
  \bvec{\mathcal{V}}(r')
  \bvec{\Phi}_0^{(r)}(r')
  \nonumber\\
  &&
  +
  \frac{2m}{\hbar^2}
  \bvec{\Phi}^{(\pm)}(r)
  \bvec{\mathcal{C}}
  \int_0^r dr'
  \bvec{\Phi}^{(r)\mathsf{T}}(r')
  \bvec{\mathcal{V}}(r')
  \bvec{\Phi}_0^{(r)}(r').
  \label{gr013mod}
\end{eqnarray}

By taking the ``{\it trace}'' of Eqs.(\ref{gr013-c1}) and (\ref{gr013-c2}),
and applying the following basic properties of the {\it trace} of the matrix
\begin{enumerate}
\item $\Tr[AB]=\Tr[BA]$
\item $\Tr[A+B]=\Tr[A]+\Tr[B]$
\item $\Tr[A^{\mathsf{T}}B]=\Tr[AB^{\mathsf{T}}]
  =\Tr[B^{\mathsf{T}}A]=\Tr[BA^{\mathsf{T}}]$,
\end{enumerate}
we can notice that Eq.(\ref{gr013-c1}) becomes
\begin{eqnarray}
  \frac{2m}{\hbar^2}
  \Tr\left[
    \bvec{\mathcal{C}}
    \bvec{\mathcal{W}^{(\pm)}}
    \right]
  =
  \Tr[\bvec{1}]
  \label{gr013-c3},
\end{eqnarray}
and Eq.(\ref{gr013-c2}) is a trivial identity. Therefore we find
\begin{eqnarray}
  \bvec{\mathcal{C}}
  =
  \frac{\hbar^2}{2m}
  \bvec{\mathcal{W}}^{(\pm)-1}
  =
  \pm 
  \frac{1}{i}
  \bvec{\mathcal{K}}
  \bvec{\mathcal{J}}^{(\pm)-1}.
  \label{C-coef}
\end{eqnarray}
By inserting Eq.(\ref{C-coef}) into Eq.(\ref{gr013mod}), finally we can derive
\begin{eqnarray}
  &&
  \bvec{\Psi}^{(\pm)}(r)
  \nonumber\\
  &&
  =
  \bvec{\Psi}^{(\pm)}_0(r)
  +
  \int_0^\infty dr'
  \bvec{\mathcal{G}}^{(\pm)}(r,r')
  \bvec{\mathcal{V}}(r')
  \bvec{\Psi}^{(\pm)}_0(r')
  \label{LSeq1}
\end{eqnarray}
with use of the definition of the Green's function Eq.(\ref{Greendef1}), 
where $\bvec{\Psi}^{(\pm)}(r)$ and $\bvec{\Psi}^{(\pm)}_0(r)$ are the
RPA and unperturbed ``scattering'' wave functions defined by
\begin{eqnarray}
  \bvec{\Psi}^{(\pm)}(r)
  &=&
  \bvec{\Phi}^{(r)}(r)
  \bvec{\mathcal{J}}^{(\pm)-1}
  \\
  \bvec{\Psi}^{(\pm)}_0(r)
  &=&
  \bvec{\Phi}^{(r)}_0(r)
  \bvec{\mathcal{J}}_0^{(\pm)-1}.
\end{eqnarray}
Since Eq.(\ref{LSeq1}) is the Lippmann-Schwinger equation, it is proved that
the RPA Green function is given by Eq.(\ref{Greendef1}).


\end{document}